\documentclass[superscriptaddress,  amsmath,amssymb, aps, twocolumn, pre]{revtex4-1}

\usepackage{graphicx} 
\usepackage{dcolumn} 
\usepackage[table,xcdraw]{xcolor}
\usepackage{graphicx}
\usepackage{textcomp}
\usepackage{braket}
\usepackage{bm}
\usepackage{multirow}
\usepackage{breqn}
\usepackage{booktabs}
\usepackage{hyperref}

\newcommand{\rua}{{\color{red}\uparrow}}

\newcommand{\rda}{{\color{red}\downarrow}}

\newcommand{\ua}{\uparrow}
\newcommand{\da}{\downarrow}
\newcommand{\majulab}{MajuLab, CNRS-UCA-SU-NUS-NTU International Joint Research Unit, Singapore}  
\newcommand{\ca}{Universit\'e C$\hat{o}$te d’Azur, CNRS, INPHYNI, Nice, France} 
\newcommand{\cqt}{Centre for Quantum Technologies, National University of Singapore, 117543 Singapore, Singapore}    
\newcommand{\nusphys}{Department of Physics, National University of Singapore, 2 Science Drive 3, Singapore 117542, Singapore} 
\newcommand{\ntuphys}{School of Physical and Mathematical Sciences, Nanyang Technological University, 637371 Singapore, Singapore}    
\newcommand{\nuscomp}{School of Computing, National University of Singapore, Singapore} 
\newcommand{\sm}{Science and Mathematics Cluster, Singapore University of Technology and Design, 8 Somapah Road, 487372 Singapore}
\newcommand{\epd}{Engineering Product Development Pillar, Singapore University of Technology and Design, 8 Somapah Road, 487372 Singapore}

\begin{document}

\title{Transfer learning for scalability of neural-network quantum states} 

\author{Remmy Zen}
\affiliation{\nuscomp} 
\author{Long My} 
\affiliation{\nuscomp}   
\author{Ryan Tan} 
\affiliation{\epd} 
\author{Fr\'ed\'eric H\'ebert} 
\affiliation{\ca} 
\author{Mario Gattobigio}        
\affiliation{\ca}  
\author{Christian Miniatura} 
\affiliation{\ca}
\affiliation{\majulab} 
\affiliation{\cqt} 
\affiliation{\nusphys} 
\affiliation{\ntuphys}  
\author{Dario Poletti} 
\email{dario\_poletti@sutd.edu.sg} 
\affiliation{\epd} 
\affiliation{\majulab}    
\affiliation{\sm}   
\author{St\'ephane Bressan} 
\affiliation{\nuscomp}

\date{\today}
\begin{abstract}
Neural-network quantum states have shown great potential for the study of many-body quantum systems. In statistical machine learning, transfer learning designates protocols reusing features of a machine learning model trained for a problem to solve a possibly related but different problem. We propose to evaluate the potential of transfer learning to improve the scalability of neural-network quantum states. We devise and present physics-inspired transfer learning protocols, reusing the features of neural-network quantum states learned for the  computation of the ground state of a small system for systems of larger sizes. We implement different protocols for restricted Boltzmann machines on general-purpose graphics processing units. This implementation alone yields a speedup over existing implementations on multi-core and distributed central processing units in comparable settings. We empirically and comparatively evaluate the efficiency (time) and effectiveness (accuracy) of different transfer learning protocols as we scale the system size in different models and different quantum phases. Namely, we consider both the transverse field Ising and Heisenberg XXZ models in one dimension, and also in two dimensions for the latter, with system sizes up to $128$ and $8 \times 8$ spins. We empirically demonstrate that some of the transfer learning protocols that we have devised can be far more effective and efficient than starting from neural-network quantum states with randomly initialized parameters.
\end{abstract}

\maketitle
\graphicspath{ {./figures/} }

\section{Introduction}

Strongly interacting quantum systems are notoriously hard to simulate because the size of their many-body Hilbert vector space grows exponentially fast with the number of particles, restricting exact diagonalization methods to few particles in practice. However, over the years, various advanced numerical methods have been developed to study these systems with an increasing degree of success. Some of the most commonly used and successful ones are quantum Monte Carlo methods~\cite{gubernatis2016,sandvik1997finite}, tensor network algorithms~\cite{orus2014practical,verstraete2004renormalization,schollwock2005density,schollwock2011} (which stemmed from the density matrix renormalization group \cite{white1992density}) and dynamical mean-field theory \cite{georges1996,Vollhardt1989, GeorgesKotliar1992}. 
Recently, a new class of techniques in the family of variational quantum Monte Carlo methods was introduced in the field and quickly came to prominence: neural-network quantum states \cite{carleo2017solving,choo2018symmetries,czischek2018quenches,deng2017quantum}. Starting with \cite{carleo2017solving}, it was shown that even simple neural networks, such as restricted Boltzmann machines, can accurately describe the ground state of a many-body quantum system, reconstruct its state and its dynamics~\cite{melko2019restricted, sarma2019review, Carleo2019review}. This is why neural-network quantum states have opened a new research direction that is widely explored today even though, for some of the most challenging systems (like frustrated two-dimensional quantum systems), more established methods can provide lower ground state energies~\cite{ChooCarleo2019}.                     

It is now possible to use open-source code libraries, such as NetKet~\cite{carleo2019netket}, to implement neural-network quantum states. NetKet is a Python framework implemented in C++ with support for the Message Passing Interface for distributed and parallel computing. 
At the same time, the recent advances in machine learning are not only due to better algorithms and packages, but also to the use of graphical processing units (GPUs), which can significantly speed up the computations at parity of hardware cost. 

A noticeable tool in the toolbox of machine learning techniques is \emph{transfer learning}~\cite{dietterich1997special}. Transfer learning proposes to use a machine learning model trained for a particular task to perform another, possibly related but different, task. Here, we use transfer learning to scale the quantum many-body system under study by transferring the optimal neural-network quantum states parameters obtained for an initial system to a similar system of larger size. We study the efficiency and effectiveness of various, physics-inspired, transfer learning protocols in different quantum phases.  By efficiency, we mean the time needed to optimize the neural-network quantum states while the effectiveness measures the accuracy of the state obtained. A transfer learning protocol combining both good efficiency and effectiveness as the system sizes grow provides good scalability. We empirically demonstrate that some of the transfer learning protocols that we have devised are far more effective and efficient than starting from a neural-network quantum states with randomly initialized parameters. Concurrently, we have ported the NetKet implementation of the neural-network quantum states to the TensorFlow machine learning platform~\cite{abadi2016tensorflow}. Consequently, our code can be readily used on general-purpose graphics processing unit. This port immediately achieves a significant performance speed-up over the NetKet code running on a cluster of commodity servers at a comparable cost of hardware. 

The remainder of this paper is structured as follows. In Sec.~\ref{sec:nnqs}, we summarize the neural-network quantum states technique. We describe the models used in Sec.~\ref{sec:experiment_setup}. In Sec.~\ref{sec:transfer_learning}, we describe the transfer learning protocols proposed for these models. In Sec.~\ref{sec:performance}, we explain how we analyze the performances of our transfer learning protocol. We first introduce the observables analyzed in Sec.~\ref{sec:evaluation}, and the neural-network quantum states implementation in Sec.~\ref{sec:implementation}. Section \ref{sec:results} is devoted to our results: Sec. \ref{sec:GPUs} details the performance of our general-purpose graphics processing unit over distributed central processing units and Sec. \ref{sec:transfer_learning_results} compares the performances of different transfer learning protocols. We briefly summarize our key results and conclude in Sec. \ref{sec:conclusions}.

\section{Neural-Network Quantum States}\label{sec:nnqs}

We summarize here how the neural-network quantum states are used to estimate the ground state of a quantum many-body system. 
This is a standard optimization problem based on the Ritz variational method~\cite{gubernatis2016}. Given a Hamiltonian $H$, the expectation value of the energy in any given state $\ket{\psi}$ is always greater than or equal to the ground state energy $E_0$, that is 

\begin{equation}
\label{eq:epsi}
E[\psi] = \frac{\braket{\psi | H | \psi}}{\braket{\psi|\psi}} \geq  E_0. 
\end{equation}
Therefore, in order to estimate the ground state energy of $H$, we start with a trial wave function $\psi(\bm{\theta})$ that depends on some parameters collectively labelled by $\bm{\theta}$ and minimize $E[\psi(\bm{\theta})]$ with respect to $\bm{\theta}$.  
\begin{figure}[t]
    \includegraphics[width=0.8\columnwidth]{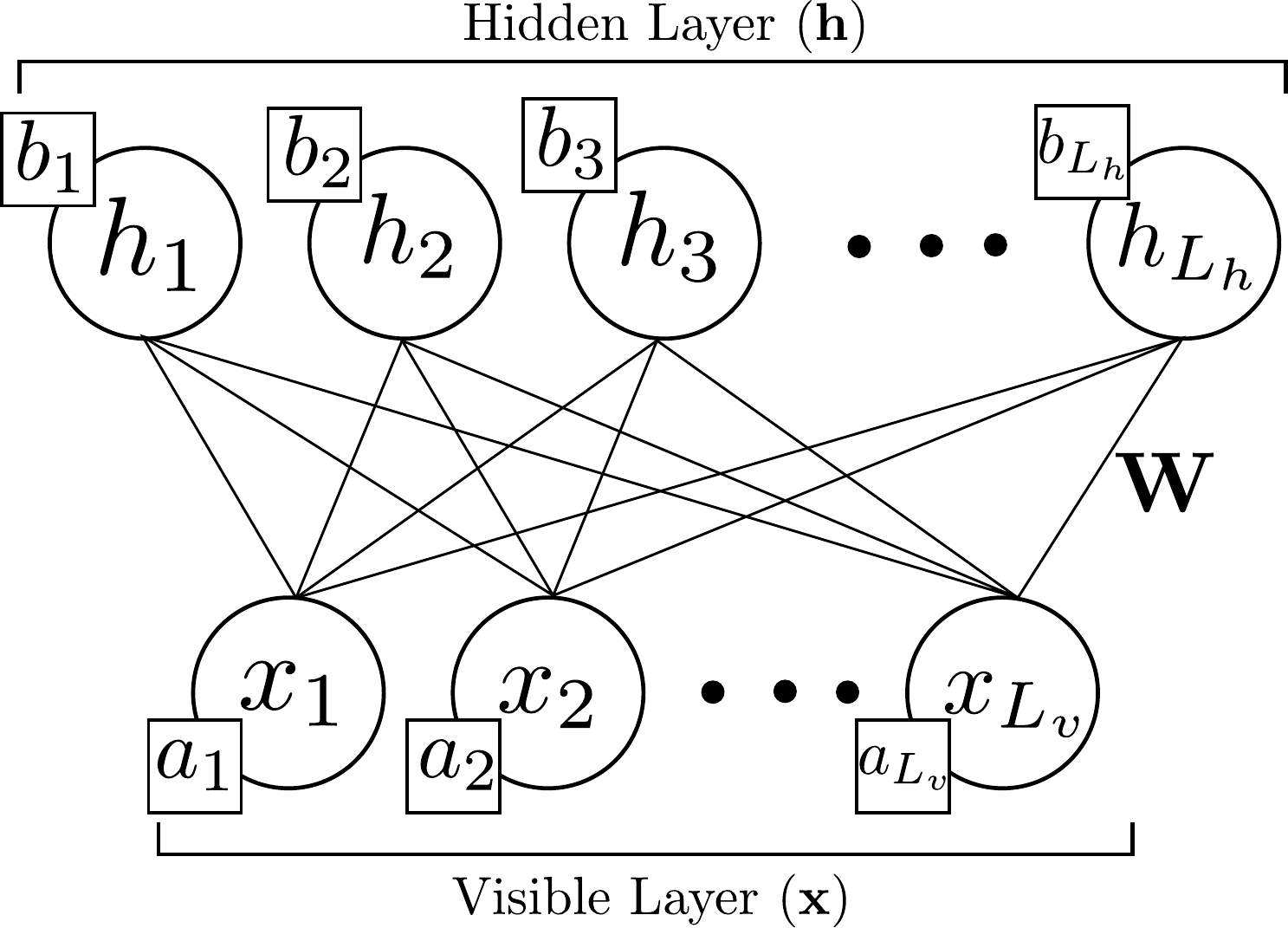}
    \caption{The structure of a restricted Boltzmann machine with $L_v$ visible nodes and $L_h$ hidden nodes. The visible layer consists of visible nodes $x_1, \dots, x_{L_v}$. The hidden layer consists of hidden nodes $h_1, \dots, h_{L_h}$. The connections between the visible and hidden layer are given by the weight matrix $\bm{W}$. The biases for the visible and hidden layers are represented by $a_1,\dots,a_{L_v}$ and $b_1,\dots,b_{L_h}$, respectively.}
    \label{fig:rbm}
\end{figure}

In this work, we use a particular type of neural network, called a restricted Boltzmann machine (RBM), to represent the wave function. It is a generative energy-based probabilistic graphical model made of two layers. The first one, called the visible layer $\bm{x}$, consists of $L_v$ nodes and is in one-to-one correspondence with the configuration space of the system. The second one, called the hidden layer $\bm{h}$, consists of an arbitrary number $L_h = \alpha L_v$ of nodes (the rational parameter $\alpha$ sets the ratio between the numbers of nodes in the two layers). 

In this work, following \cite{carleo2017solving}, we concentrate on spin 1/2 models, such as the transverse field Ising or the anisotropic Heisenberg models.  
To describe such systems, binary values $\{+1, -1\}$, corresponding to the projection of the spins along the $z$ axis, are then assigned to the visible nodes variables $x_j$, which spans the entire Hilbert space of the system. The same binary values are used for the hidden nodes $h_i$.

As shown in Fig.\ref{fig:rbm}, the structure of a restricted Boltzmann machine forms a bipartite graph in which the hidden and visible nodes are associated with a set of weights $\bm{W}=\{W_{ji}\}$ where the first and second index in the matrix label, respectively, the nodes of the visible and hidden layers (e.g. $i \in \{1,\dots,L_h\}$ and  $j \in \{1,\dots,L_v\}$). The neural network has also a visible bias vector $\bm{a}=\{a_{j}\}$ and a hidden bias vector $\bm{b}=\{b_{i}\}$ which couple to each node in their respective layer. 
In our work, all the elements of $\bm{W}$, $\bm{a}$ and $\bm{b}$ are taken as real numbers, which is sufficient to represent the ground states of the systems we are studying. Using the restricted Boltzmann machine, the joint probability distribution of a configuration of the visible layer $\bm{x}$, and a configuration of the hidden layer $\bm{h}$, for given $\bm{W}$, $\bm{a}$ and $\bm{b}$, is given by 
\begin{eqnarray}
p^{}_{RBM}(\bm{x}, \bm{h}; \bm{W}, \bm{a}, \bm{b}) &=& \frac{\exp\left(\bm{a}\!\cdot\!\bm{x} + \bm{b}\!\cdot\!\bm{h} + \bm{x}\!\cdot\! \bm{W}\bm{h}\right)}{Z} 
\label{eq:prbm}
\end{eqnarray}
where $\bm{W}\bm{h}$ is the vector obtained by applying the matrix $\bm{W}$ onto $\bm{h}$ and $Z$ normalizes the probability so that the sum over all possible combinations of $\bm{x}$ and $\bm{h}$ is $1$.

From Eq.~(\ref{eq:prbm}), by integrating out the hidden layer, we can compute the marginal distribution of the visible layer $p^m_{RBM}(\bm{x}; \bm{W}, \bm{a}, \bm{b})$:  
\begin{eqnarray}
\label{eq:marg_prbm}
p^{m}_{RBM}(\bm{x}; \bm{W}, \bm{a}, \bm{b}) &=& \frac{1}{Z} \exp \left( \bm{a}\cdot\bm{x} \right) \\&\times&
\prod_{i}2\cosh{\left(\sum_{j} x_j W_{ji} + b_i\right).}\nonumber
\end{eqnarray}
As in \cite{carleo2017solving}, we use the marginal distribution provided by the restricted Boltzmann machine to represent the probability of a given configuration $\bm{x}$ of the ground state with parameters $\bm{\theta}=\{\bm{W},\bm{a},\bm{b}\}$. Using the bra-ket quantum notation, the (normalized to $1$) trial wave function $|\psi(\bm{\theta})\rangle = \sum_{\bm{x}} \psi(\bm{x};\bm{\theta}) \, |\bm{x}\rangle$ returned by the restricted Boltzmann machine is thus defined by $ \psi(\bm{x};\bm{\theta})=\sqrt{p_{\psi}(\bm{x};\bm{\theta})} = \langle \bm{x} | \psi(\theta)\rangle $ where:

\begin{equation}
\label{eq:conn}
p_{\psi}(\bm{x}; \bm{\theta}) = p^{m}_{RBM}(\bm{x}; \bm{W}, \bm{a}, \bm{b}).
\end{equation}
Having real and positive coefficients $\psi(\bm{x};\bm{\theta})$, the trial wave function $|\psi(\bm{\theta})\rangle$ is well adapted to the models we intend to study because their respective ground states can be taken positive.

From Eqs.~(\ref{eq:epsi},\ref{eq:prbm},\ref{eq:conn}), the problem of estimating the ground state energy becomes the problem of minimizing the function $E[\psi(\bm{\theta})] = E(\bm{\theta})$ given by 
\begin{align}
    E(\bm{\theta}) &=  \frac{\langle \psi(\bm{\theta})| H | \psi(\bm{\theta})\rangle} {\langle \psi(\bm{\theta})|\psi(\bm{\theta})\rangle} 
    = \sum_{\bm{x}} p_{\psi}(\bm{x};\bm{\theta})E_{\mathrm{loc}}(\bm{x};\bm{\theta}).  \label{eq:loss}     
\end{align}
Using the completeness relation $\sum_{\bm{x}} |\bm{x}\rangle\langle\bm{x}| = \openone$, the local energy $E_{\mathrm{loc}}$ is given by:
\begin{align}
    E_{\mathrm{loc}}(\bm{x}; \bm{\theta}) = \sum_{\bm{x'}} \braket{\bm{x} \mid H \mid \bm{x'}} \frac{\psi(\bm{x}';\bm{\theta})}{\psi(\bm{x};\bm{\theta})}. 
    \label{eq:eloc} 
\end{align}

The minimization of $E(\bm{\theta})$ uses a stochastic gradient descent algorithm to iteratively update the parameters $\bm{\theta}$. The value of $E(\bm{\theta})$ and its gradients are calculated by taking samples from $p_{\psi}(\bm{x}; \bm{\theta})$. 

The sampling procedures rely on Gibbs sampling and the Metropolis-Hastings algorithm. 
For Gibbs sampling, starting from an initial visible configuration $\bm{x}$, a hidden configuration is generated by sampling from the conditional probability $p(\bm{h}|\bm{x})$ given by  
\begin{equation}
    \label{eq:hgivenx}
    p(\bm{x}|\bm{h}) = \prod_i \mathrm{sigmoid}\left[2\left(\sum_{j}W_{ji}x_j + b_i\right)h_i\right], 
\end{equation}
then from this hidden configuration, a new visible configuration is generated by sampling from the conditional probability $p(\bm{h}|\bm{x})$ given by 
\begin{equation}
    \label{eq:xgivenh}
    p(\bm{h}|\bm{x}) = \prod_j \mathrm{sigmoid}\left[2\left(\sum_{i}W_{ji}h_i + a_j\right)x_i\right]. 
\end{equation}
For the Metropolis-Hastings algorithm, starting from an initial visible configuration $\bm{x}$, we get a new visible configuration $\bm{x'}$ through an arbitrary strategy, and decide to accept or reject it depending on the relative probability $p^{m}_{RBM}(\bm{x'}; \bm{\theta}) / p^{m}_{RBM}(\bm{x}; \bm{\theta})$. Gibbs sampling is typically much faster than the Metropolis-Hastings sampling algorithm, as auto-correlation times in the sampling are longer in the latter case. However, the Metropolis-Hastings algorithm is much more suitable to impose constraints on the possible samples. For example, for cases where the states are restricted to a fixed total magnetization along the $z$ axis (which means that the total number of up and down spins is fixed), a sampling algorithm obeying such criterion can be readily implemented in the Metropolis-Hastings algorithm by using spin exchanges to propose new configurations.

\section{Models analyzed}
\label{sec:experiment_setup}

We conduct numerical experiments on two different quantum models, namely the transverse field Ising and Heisenberg XXZ models. 
Hereafter, we refer to the former as the Ising model and to the latter as the Heisenberg model. They are described, respectively, by the Hamiltonians 
\begin{align}
H_{I} = -J_I \sum_{\langle l,m\rangle} \sigma^z_l\sigma^z_m - h \sum_l \sigma^x_l \label{eq:H_I}     
\end{align}
and 
\begin{align}
H_{XXZ} = - J_{XXZ} \sum_{\langle l,m\rangle} \left(\sigma^x_l\sigma^x_m + \sigma^y_l\sigma^y_m  + \Delta \sigma^z_l\sigma^z_m\right) , 
\label{eq:H_H}      
\end{align} 
where the $\sigma^{x/y/z}_l$ are operators acting on site $l$ and corresponding to the respective Pauli matrices. For the Ising model, we use the parameters $J_I$ for the spin coupling strength and $h$ for the transverse field, while for the Heisenberg model we use $J_{XXZ}$ for the in-plane spin coupling strength and $\Delta$ for the dimensionless anisotropy factor between the $xy$ plane and $z$ axis coupling strengths. For both models, we consider open boundary conditions, for which simplifications of the restricted Boltzmann machine leveraging on translation invariance~\cite{carleo2017solving,choo2018symmetries} cannot be used. 

For the Ising model, the spin coupling $J_I$ favors a ferromagnetic (F) ground state when positive and an antiferromagnetic (AF) one when negative. At the same time, a strong magnetic field $h$ favors spins that are aligned paramagnetically, pointing along the $x$ axis in the positive ($h> 0$) or negative ($h<0$) direction. In the one-dimensional model, a quantum phase transition occurs at $|J_I/h|=1$. The ground state being ferromagnetic along the $z$ axis for $J_I>|h|$, antiferromagnetic along $z$ for $J_I<-|h|$ and paramagnetic (PM) otherwise.

For the Heisenberg model, there is an exact mapping between Hamiltonians with parameters $(J_{XXZ},\Delta)$ and $(-J_{XXZ},-\Delta)$ and we will then keep $J_{XXZ}=1$. In one-dimension, for $J_{XXZ}>0$, the ground state can be in three different phases: ferromagnetic along $z$ for $\Delta>1$, showing no magnetic order along $z$ for $|\Delta|<1$  (where the model orders ferromagnetically in the $xy$-plane) and antiferromagnetic along $z$ for $\Delta<-1$. When $\Delta=1$, the model is isotropic.

An important difference between the Ising and Heisenberg models is that the latter preserves the total spin along $z$, which commutes with $H_{XXZ}$ and is then a conserved quantity, while, for the former, there is no such conservation. In the following, we restrict our study of the Heisenberg model to the sector where the total magnetization along $z$ is zero. Then, if the system becomes ferromagnetic along the $z$ axis, its lowest-energy configuration will correspond to a symmetric superposition of two states each composed of one spin-up domain and one spin-down domain of equal sizes, separated by a single domain wall exactly located in the middle of the system. 

\section{Transfer Learning Protocols} \label{sec:transfer_learning}

\begin{figure} 
\includegraphics[width=\columnwidth]{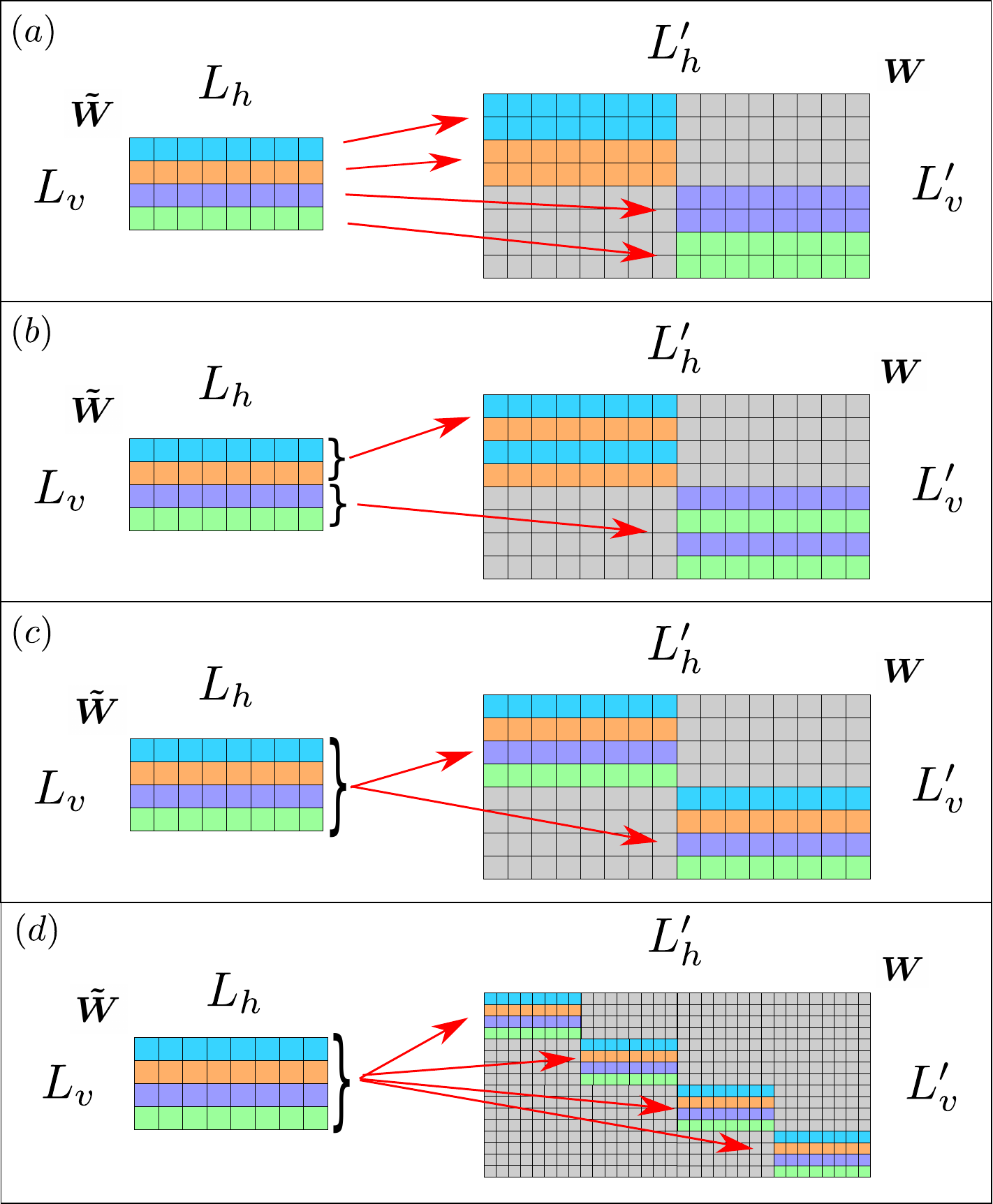} 
\caption{(Color online) Schematic representation of the different transfer learning protocols used to scale up one-dimensional systems. $L_v$, $L_h$, and $L'_v$ and $L'_h$ correspond to the number of visible and hidden nodes for the base and target networks, respectively. The different panels show how to construct the target weight matrix $\bm{W}$ from the base weight matrix $\bm{\tilde{W}}$ by replicating the colored rows and filling the grey cells with randoms entries. Panels (a), (b), and (c): $(1,2)-$tiling, $(2,2)-$tiling, and $(L,2)-$tiling respectively. Here the scaling factor is $2$. Panel (d) shows the $(L,4)-$tiling used when scaling up the network by a factor $4$; See text for a detailed explanation.} \label{fig:kp_tiling}   
\end{figure}

In machine learning, transfer learning refers to the reuse of the features of a machine learning model learned for the resolution of a problem to initialize a machine learning model used to solve another problem~\cite{dietterich1997special}. The authors of~\cite{yosinski2014transferable} discuss the transferability of features between deep neural networks with the same architecture, i.e. the same nodes and connections. 

In our case, a many-body system is simulated by a restricted Boltzmann machine. We consider the transfer of parameters to networks having, however, different architectures. With transfer learning, we aim to solve a larger system problem using the solution of a smaller one, although there are applications for the reverse. The transfer learning approach is as follows: First, we train a network using randomly initialized parameters to minimize the energy of a system. Following the terminology in~\cite{yosinski2014transferable}, we refer to this network as the \emph{base network}. Secondly, after the base network has been trained, we copy the parameters of the network to another larger system, that is called the \emph{target network}. Thirdly, we train the target network using the transferred parameters as the initialized parameters of the network. We call this process the \emph{fine-tuning} of the network. Lastly, we evaluate the transfer learning protocol by comparing the performance of the network trained with the transfer learning protocol and the network trained from scratch with randomly initialized parameters. The former is commonly called the \emph{hot-start} network while the latter is called the \emph{cold-start} network. 

In this Paper, we devise different transfer learning protocols and comparatively evaluate their efficiency and effectiveness.  To evaluate the efficiency, we compare the times needed to reach an energy minimum. To evaluate the effectiveness, we probe the accuracy of the wave function. Starting from a one dimensional system of $L_v$ spins, we increase the number of spins to $L_v'=2L_v,\;3L_v,\;4L_v,\dots$ at each iteration. We denote the weights of the base network by $\tilde{W}_{ji}$ and the weights of the target network by $W_{ji}$. The transfer learning rule, or protocol, is then specified by the mapping $\tilde{W}_{ji} \to W_{ji}$. Mathematically, there is a large number of possible ways to transfer the weights from a smaller system to a larger one, and it would not be possible, nor meaningful, to consider all of them.  Instead, we consider a transfer learning protocol which we refer to as \emph{$(k,p)-$tiling} where groups of $k$ weights calculated for a system at a certain size are repeated $p$ times to initialize the weights for a larger system.  
Even if our transfer learning protocols only focus on weight transfer, our simulations show that, after the fine-tuning process, the bias values are generally negligible. We have further checked that simulations with biases set to zero and with variable biases yield very close results. 

The different tilings that we used are detailed below and depicted in Fig.\ref{fig:kp_tiling}. Do note that each tiling is not just performed over the same hidden nodes but is distributed in equal parts over portions of the hidden nodes. By doing so, all the hidden nodes of the target network retain some knowledge acquired by the base network. This was found to help the optimization process of the target network.  

\begin{itemize}
\item {$(1,2)-$tiling} (see Fig.\ref{fig:kp_tiling}(a)). 
We focus on doubling the system ($p = 2$). The weights of the target network are initialized with $W_{2j-1,i}=W_{2j,i}=\tilde{W}_{j,i}$ for $j\in[1,L_v/2]$, $W_{2j-1,i+L_h}=W_{2j,i+L_h}=\tilde{W}_{j,i}$ for $j\in[L_v/2+1,L_v]$, and a random value for all the other terms.  This protocol is expected to work well for paramagnetic or ferromagnetic phases, however it may not be the ideal transfer learning for a system in the antiferromagnetic phase. This tiling indeed favors situations in which each site is equivalent, e.g. considering the state $|\uparrow\uparrow\uparrow\uparrow\rangle$, the doubling will give a bias towards the state $|\uparrow\rua\uparrow\rua\uparrow\rua\uparrow\rua\rangle$ (color online, spins in red are ``copied'' from the base network). This could be seen as favoring ferromagnetic correlations. We see a similar effect in a ferromagnetic state where the total magnetization along $z$ is fixed to zero ($|\uparrow\uparrow\downarrow\downarrow\rangle$ to $|\uparrow\rua\uparrow\rua\downarrow\rda\downarrow\rda\rangle$). If however the base network represents an antiferromagnetic state $|\uparrow\downarrow\uparrow\downarrow\rangle$, then the initialization of the target network would bias towards the state $|\uparrow\rua\downarrow\rda\uparrow\rua\downarrow\rda\rangle$ which has much weaker antiferromagnetic correlations.\\  
\item {$(2,2)-$tiling} (see Fig.\ref{fig:kp_tiling}(b)). This protocol is similar to the $(1,2)-$tiling just described, but instead of doubling a single spin, we double a pair of them since $k = 2$.  We use $W_{4j-3,i}=W_{4j-1,i}=\tilde{W}_{2j-1,i}$ and $W_{4j-2,i}=W_{4j,i}=\tilde{W}_{2j,i}$ for $j\in[1,L_v/4]$, while $W_{4j-3,i+L_h}=W_{4j-1,i+L_h}=\tilde{W}_{2j-1,i}$ and $W_{4j-2,i+L_h}=W_{4j,i+L_h}=\tilde{W}_{2j,i}$ for $j\in[L_v/4+1,L_v/2]$, and a random value for all the other terms.  
This protocol is expected to work better for antiferromagnetic phases compared to the $(1,2)-$tiling protocol, as unit cells of two sites are copied, while still being effective for ferromagnetic phases. A state $|\ua\da\ua\da\rangle$ would give a bias towards the state $| \ua\da\rua\rda\ua\da\rua\rda  \rangle$, which preserves antiferromagnetic correlations. Similarly in the ferromagnetic phase, a state $|\ua\ua\ua\ua\rangle$ would bias towards the state $| \ua\ua\rua\rua\ua\ua\rua\rua \rangle$ and, in the case of a zero magnetization ferromagnetic state, $|\uparrow\uparrow\downarrow\downarrow\rangle$ would bias towards the state $| \ua\ua\rua\rua\da\da\rda\rda  \rangle$, which preserves ferromagnetic correlation.\\    
\item {$(L,p)-$tiling} (see Fig.\ref{fig:kp_tiling}(c) for the $(L,2)$-tiling and Fig.\ref{fig:kp_tiling}(d) for the $(L,4)$-tiling). By setting $k = L_v$, we transfer all the base network weights $\tilde{W}_{j,i}$ on the first $L_v$ visible nodes, and then repeat them for the other $(p-1)L_v$ half on the visible layer nodes but coupled to the other hidden nodes. We stress here that the symbol $L$ denotes the whole visible nodes, $L = L_v$. More precisely, we match $W_{j+\eta L_v,i+\eta L_h}=\tilde{W}_{j,i}$ for $j\in[1,L_v]$, $i\in[1,L_h]$ and $\eta\in[0,p-1]$. This protocol could be favorable in the antiferromagnetic phase because a state $|\ua\da\ua\da\rangle$ would give a bias towards the state $| \ua\da\ua\da\rua\rda\rua\rda  \rangle$ and in the ferromagnetic phase because a state $|\ua\ua\ua\ua\rangle$ would give a bias towards the state $| \ua\ua\ua\ua\rua\rua\rua\rua  \rangle$. However, for a ferromagnetic state where the sum of spins is fixed to zero, the state $|\ua\ua\da\da\rangle$ would give some bias towards $| \ua\ua\da\da\rua\rua\rda\rda \rangle$ which has weaker ferromagnetic correlations.\\     
\end{itemize} 

Examples of the $(1,2)$, $(2,2)$, and $(L,2)-$tiling protocols are depicted in Fig.~\ref{fig:kp_tiling} (a,b,c), in the case where the transfer is from $L_v = 4$ to  $L'_v = 8$  with $\alpha = 2$.  Figure \ref{fig:kp_tiling}(d) shows the $(L,4)-$tiling from $L_v = 4$ to  $L'_v = 16$. 

A natural generalization of $(k,p)-$tilings to lattices in higher dimensions $d>1$ are $(\bm{k},\bm{p})$-tilings where $\bm{k}$ and $\bm{p}$ are $d$-dimensional vectors. In this case, each couple $(k_a, p_a)$ ($1\leq a\leq d$) tells how many groups of weights computed for a certain system size to consider and how many times to repeat them to initialize the weights for a larger system size along the $a-$th axis. Hereafter, for simplicity, we consider the case of the isotropic transfer learning protocol where $k_a =k$ and $p_a = p$  for all $a$. Hence we will, once again, just refer to this transfer protocol as the $(k,p)-$tiling. The above mentioned protocols may not be exhaustive, but as we will see later, they already give an interesting insight into the efficiency and effectiveness of the transfer learning method.

\section{Performance Evaluation}\label{sec:performance}

In this section, we describe the quantities we use to evaluate the performance in Sec.~\ref{sec:evaluation} and the implementation details of the neural-network quantum states and its minimization procedure in Sec.~\ref{sec:implementation}.  

To explore the different phases of the one-dimensional Ising and Heisenberg models, we fixed $h=1$ and $J_{XXZ}=1$ (the latter ensures that the ground state can be taken as positive) and we have varied $J_I$ and $\Delta$, concentrating on three values, each corresponding to a different phase. 

For the one-dimensional Ising model, we used $J_I = 2$, in the ferromagnetic phase (later referred to as Ising F), $J_I=-2$, in the antiferromagnetic one (Ising AF), and $J_I=0.5$, in the paramagnetic phase (Ising PM). For the one-dimensional Heisenberg model, we studied the cases $\Delta=2$, ferromagnetic along the $z$ axis (Heis F), $\Delta=-2$, antiferromagnetic along the $z$ axis (Heis AF) and $\Delta=-0.5$, ferromagnetic in the $xy$ plane but showing no order along the $z$ axis (Heis XY).

For the two-dimensional Heisenberg model, we concentrated on the case with $\Delta=-1$ and $J_{XXZ}=1$ which is, through the previously mentioned mapping, equivalent to the $\Delta=1$, $J_{XXZ}=-1$ isotropic antiferromagnetic Heisenberg model.

In the following, we consider chains containing $L_v=\{4,8,16,32,64,128\}$ spins for the one-dimensional Ising and Heisenberg models and $L_v=\{2\times2,4\times4,8\times8\}$ square lattices for the two-dimensional Heisenberg model. As the minimization procedures that we use to compute the ground state have random components, each estimate of an observable is an average over $20$ realizations of the same calculation.

\subsection{Evaluation methods}
\label{sec:evaluation}

We evaluate both the efficiency and the effectiveness of the proposed transfer learning protocols and compare the results to other readily available methods. 
For efficiency, we refer to the time needed to reach the stopping criterion of the minimization (see below for a description of this criterion). We compare the time needed for the cold-start with the time required by the hot-start plus the time of the previous iterations. For instance, for a one-dimensional system with $L_v=32$, we compare the time required by a cold-start for this size to the time needed by a cold-start for $L_v=4$ plus the time for hot-starts at $L_v=8$, $L_v=16$ and finally $L_v=32$.   

For effectiveness, we refer to the quality of the representation of the ground state by our ansatz. 
As energy alone might not be a sufficient indicator of the quality of the ground state, we also consider 
the spin-spin correlator between two spins $l$ and $m$, $C^z_{l,m} =  \braket{\sigma_l^z \sigma_m^z}$. This is computed by 
\begin{align}
\label{eq:cz}
C^z_{l,m} =  \sum_{\bm{x}} p(\bm{x})  \langle \bm{x} | \sigma_{l}^z \sigma_{m}^z | \bm{x} \rangle, 
\end{align} 
which is evaluated with Monte Carlo sampling over the possible spin configurations $\bm{x}$. 
From Eq.~(\ref{eq:cz}), it is possible to define correlators that are useful to identify ferromagnetic order, $C^F_d$, or anti-ferromagnetic order, $C^A_d$, which are respectively 
\begin{align}\label{eq:cF} 
C^F_d=\frac{1}{d-1} \sum_{l=2}^d C^z_{1,l}
\end{align}
and 
\begin{align}\label{eq:cA} 
C^A_d=\frac{1}{d-1}  \sum_{l=2}^d (-1)^{l-1} C^z_{1,l},  
\end{align}
where $d$ is a range of distances between the two spins that we consider, for instance, half the length of the spin chain. As these correlators are specific for antiferromagnetic and ferromagnetic orders along the $z$ axis, we do not evaluate these correlators in phases that do not show a magnetic order along $z$.     

The behavior of $C^F_d$ is different in the ferromagnetic phases of the Ising and Heisenberg models. For the Ising model, we can expect 
$C^F_d$ to be almost constant for all $d$. On the contrary, for the Heisenberg model, as we work at fixed zero magnetization, $C^F_d$ should be almost constant for $d<L_v/2$ and should then decrease towards zero for $d>L_v/2$ (or vice versa) as the system is separated into two ferromagnetic domains with opposite spins.

For one-dimensional Ising and Heisenberg model, the results from neural-network quantum states calculations for ground state energy and correlations are compared to accurate matrix product states simulations~\cite{schollwock2011}. For the matrix product states simulations, we use both a non-number conserving code for the Ising model and a number conserving code for the Heisenberg model, with a bond dimension $D$ up to $1000$ for most of the computations. For the two-dimensional Heisenberg model, with $J_{XXZ}=1$,$\Delta=-1$, the ground state energy is compared to quantum Monte Carlo results~\cite{sandvik1997finite} with $\beta= 50.0$, $20$ bins and $100,000$ steps per bin. We use the mean relative error as the effectiveness metric of the ground state energy and the correlations. Furthermore, we plot the ferromagnetic correlator $C^F_d$ and the antiferromagnetic correlator $C^A_d$ for a different range of distances $d$ to evaluate the correlations qualitatively.

To quantify how much the parameters of the restricted Boltzmann machine change during an optimization from a hot start, we measure the \emph{transfer distance} $\mathcal{D}$ between the initial weights, denoted as $W^{(init)}$, of the target network after transferring from the base network, and the weights after fine-tuning the target network denoted as $W^{(final)}$. This is defined by the mean absolute difference between $W^{(init)}$ and $W^{(final)}$, that is 
\begin{align}
\label{eq:mad}
\mathcal{D}(W^{(init)}, W^{(final)}) =  \sum_{j}^{L_v}\sum_i^{L_h}\frac{ |W^{(init)}_{i,j} - W^{(final)}_{i,j}|}{L_v L_h} .  
\end{align} 
A small $\mathcal{D}$ means that the transfer gives already very good parameters for the network, and little change is needed, while a large $\mathcal{D}$ implies a less adequate transfer.

\subsection{Details of the neural-network quantum states implementation} \label{sec:implementation}      

The actual implementation of the neural-network quantum states and its minimization relies on several parameters and steps which we here present in greater detail. The value of the parameters presented here is determined from the literature or with a grid search.

The optimization process is done iteratively. In each iteration, we take $10^4$ samples to evaluate the energy $E(\bm{\theta})$ and its gradients using the current parameters of the restricted Boltzmann machine; then we update the parameters of the restricted Boltzmann machine using a gradient descent algorithm, for which, in our case, we choose adaptive learning rate strategies RMSProp~\cite{hinton2012neural} and Adam~\cite{kingma2014adam}. We set the initial learning rate to $0.001$. For the Ising model, the samples are obtained by a single iteration Gibbs sampling. For the Heisenberg model, we instead use a Metropolis-Hastings algorithm with exchange strategy, for which two random spins are exchanged to obtain a new configuration while keeping the total number of spins up and of spin down invariant. This strategy is used in the Heisenberg model to conserve the total spin magnetization. The number of thermalization steps is set to $10\%$ of the number of Monte Carlo samples.

For the restricted Boltzmann machine, based on our exploratory simulations, we found that $\alpha=2$  is a good compromise considering the efficiency and effectiveness trade-off.  
During transfer learning, or for cold-starts, we set some parameters of the restricted Boltzmann machine randomly. For the weight matrix $\bm{W}$, we sample the weight from a normal distribution $\mathcal{N}(0, 0.01)$ with mean $0$ and standard deviation $0.01$, while the elements of both biases $\bm{a}$ and $\bm{b}$ are initialized to zero based on~\cite{hinton2012practical} and our exploratory experiments. 

To halt the simulations, we use a dynamic stopping criterion based on the zero-variance principle in quantum Monte Carlo~\cite{gubernatis2016}. We compute the standard deviation of $E_{\mathrm{loc}}(\bm{x})$: 
\begin{align}
\sigma_{E_{\mathrm{loc}}} = \sqrt{ \sum_{\bm{x}} \left(E_{\mathrm{loc}}(\bm{x},\bm{\theta}) - E(\bm{\theta})\right)^2 } /\sqrt{N}   \label{eq:stop_criterion} 
\end{align}
where $N$ is the number of configurations $\bm{x}$ considered, while $E(\bm{\theta})$ and $E_{\mathrm{loc}}(\bm{x},\bm{\theta})$ are given, respectively, in Eq.~(\ref{eq:loss}) and Eq.~(\ref{eq:eloc}),
and we stop the simulation when its ratio with the average energy, $\sigma_{E_{\mathrm{loc}}}/E(\bm{\theta)}$, is lower than a threshold value $\varepsilon_\sigma$. If $\sigma_{E_{\mathrm{loc}}}/E(\bm{\theta)}$ does not reach a value smaller than $\varepsilon_\sigma$, we stop the simulation after $\varepsilon_{epoch}$ epochs. In our evaluations, we take $\varepsilon_\sigma=0.005$ and $\varepsilon_{epoch} = 30000$. 

This stopping criterion is motivated by the fact that, if the neural-network quantum states represents any eigenstate of the Hamiltonian, then $E_{\mathrm{loc}}(\bm{x},\bm{\theta})$  returns the energy of the eigenstate for any configuration $\bm{x}$. Combining this with the fact that the algorithm searches for the lowest energy, if $\sigma_{E_{\mathrm{loc}}}$ is very small then the neural-network quantum states should be a good representation of the ground state. To compute the observables, we sample over a certain number of configurations $\bm{x}$ taken from $p_{\psi}(\bm{x},\bm{\theta})$ with 500 steps and 5000 steps for Gibbs sampling and Metropolis-Hastings algorithm, respectively. These values were determined by looking at the auto-correlation times in our exploratory experiments.

\section{Results} \label{sec:results} 
\subsection{Efficiency of graphics processing unit}\label{sec:GPUs}

\begin{figure}[t]
    \includegraphics[width=0.85\columnwidth]{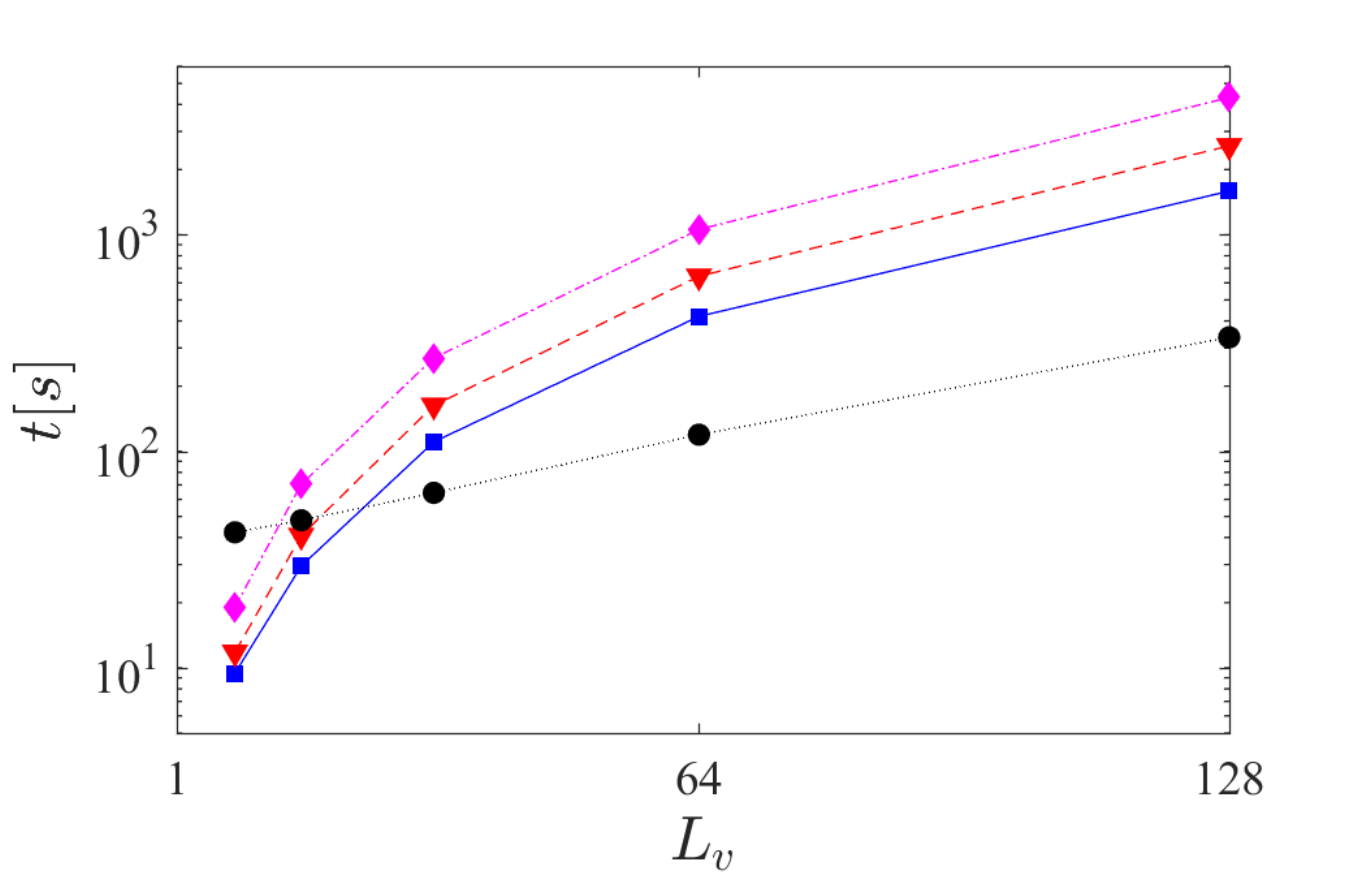}
    \caption{(Color online) Computation time ($t[s]$) as a function of the number of spins ($L_v$) for various settings on one-dimensional Ising model with $J_I=1$. The purple dash-dotted line with diamonds shows the time using four central processing units with NetKet. The red dashed line with triangles shows the time using eight central processing units with NetKet. The solid blue line with squares shows the time using sixteen central processing units  with NetKet. The black dotted line with circles shows the time using a graphics processing unit with TensorFlow.}
    \label{fig:cpugpu-1d}
\end{figure}

We implement the neural-network quantum states code with the machine learning platform TensorFlow~\cite{abadi2016tensorflow}. TensorFlow allows us to deploy and run the code on a general-purpose graphics processing unit server, which we then also compare to a parallel central processing units implementation, for which we use the NetKet library~\footnote{Here we have compared both of the codes with the same set of parameters using restricted Boltzmann machine and sampling with the Metropolis-Hastings algorithm that flips a random spin. One should notice that the NetKet implementation uses a slightly different version of the RBM than the one presented here, where the wave function is directly given by $\psi(\bm{x},\bm{\theta})=p_{RBM}^m(\bm{x};\bm{\theta})$. For a fair comparison of the performance of both codes, we used this definition instead of the one presented in Sec.~\ref{sec:nnqs} for these tests. The rest of the results were obtained using the method presented in Sec.~\ref{sec:nnqs} which allows the use of the Gibbs sampling.}.  
For the NetKet library, all experiments in this subsection run on machine with Ubuntu 16.04 operating systems with two 1.8Ghz Intel\textsuperscript{\textcopyright} Xeon\textsuperscript{\textcopyright} Silver 4108 processors and equipped with 128GB DDR4 memory. For the TensorFlow implementation, at a comparable hardware purchase price~\footnote{As a reference, at the time of writing, the cost of each processor and cost of the graphics processing units is around 500 USD.},  all experiments in this subsection run on NVIDIA GeForce GTX 1080 graphics processing unit with 2560 CUDA cores and 8GB memory. We implement the program using Python 2.7.15. 

The efficiency of our TensorFlow implementation on a general-purpose graphics processing unit server is higher than that of the original NetKet implementation on a multi-core server beyond a sufficiently large system size. This is illustrated, for reference only, in Fig.~\ref{fig:cpugpu-1d}. In this case, we have considered a one-dimensional Ising model with $8$ to $128$ spins at the quantum phase transition point where $J_I=1.0$. For a system of $128$ spins, the TensorFlow implementation runs about $5$ times faster than the NetKet implementation.

\subsection{Evaluation of transfer learning protocols}\label{sec:transfer_learning_results}    

\begin{figure}[t]
    \includegraphics[width=\columnwidth]{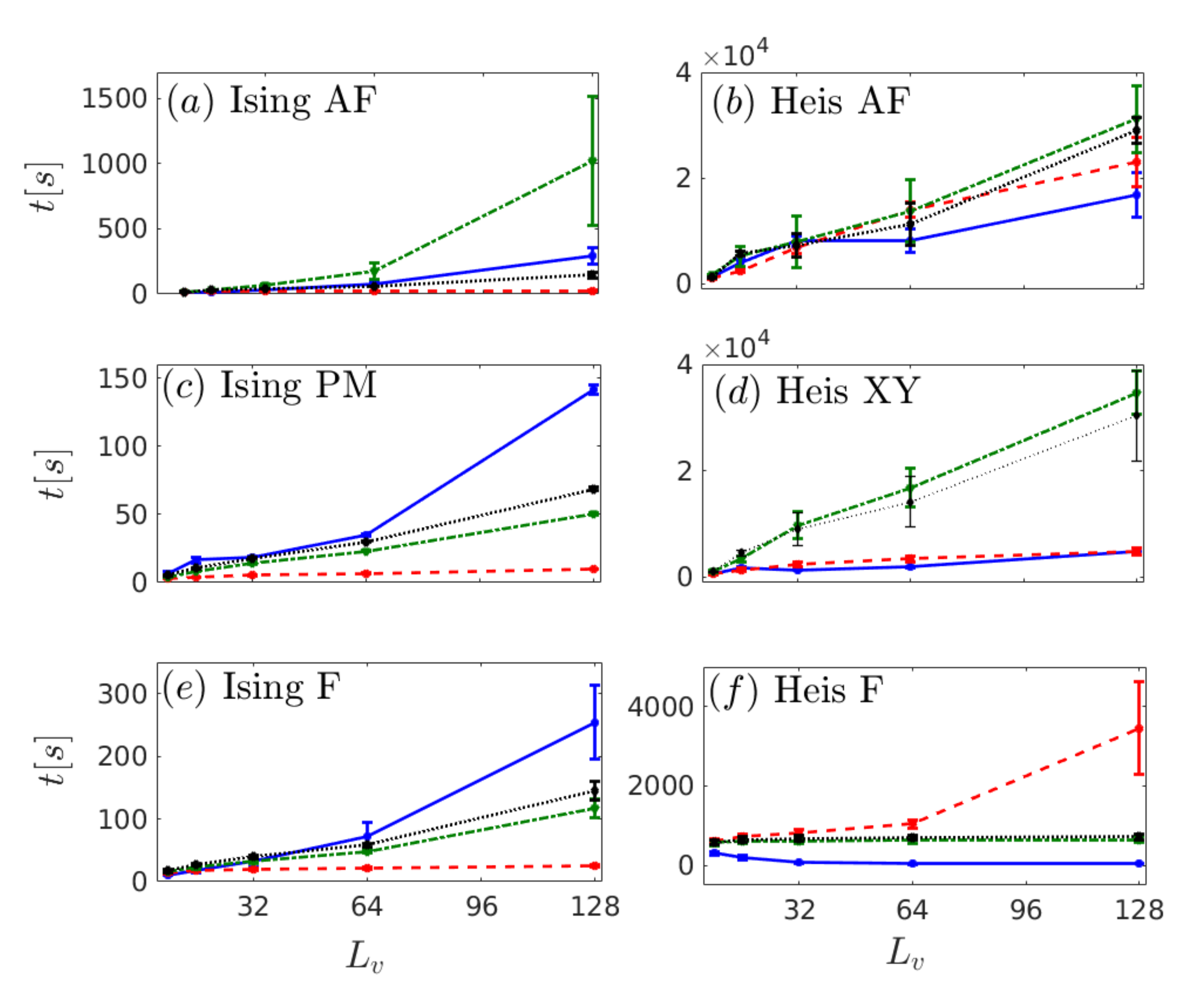}
    \caption{(Color online) The efficiency of different transfer learning protocols. Each panel shows the simulation time in seconds $t[s]$ until the stopping criterion is reached as a function of the number of spins ($L_v$). The left (a,c,e) and right (b,d,f) columns panel corresponds to Ising and Heisenberg models, respectively. The different rows correspond to the different parameters of the Hamiltonian ($J_I$ for the Ising model and $\Delta$ for the Heisenberg model) taking on values -2, $\pm$0.5 and 2, respectively. The blue solid line with squares shows the time for the cold-start. The red dashed line with circles shows the time for the $(L,2)-$tiling protocol, the black dotted line with diamonds the time for the $(2,2)-$tiling protocol, and the green dash-dotted line with triangles the time for $(1,2)-$tiling protocol. The error bars represent the interval of $\pm 1$ standard deviation. 
    In all figures, we take $h=1$ (for Ising) and $J_{XXZ}=1$ (for Heisenberg).     }
    \label{fig:fig4}
\end{figure}

All experiments in this subsection run on an NVIDIA DGX-1 server equipped with NVIDIA Tesla V100 graphics processing units with 640 tensor cores, 5120 CUDA cores, and 16GB memory. We first compare the efficiency of different transfer learning protocol and cold-start of Ising and Heisenberg model for three different Hamiltonian parameters in Fig.~\ref{fig:fig4}. In particular in Fig.\ref{fig:fig4} (a,c,e) we show results for the Ising model, respectively for $J_I=-2,\;0.5$, and $2$, while in Fig.\ref{fig:fig4} (b,d,f) we show results for the Heisenberg model, respectively with $\Delta=-2,\;-0.5$ and $2$. The error bars represent the interval of $\pm 1$ standard deviation from $20$ realizations. The error bars are used to evaluate the robustness of the transfer learning protocol, e.g. a smaller error bars corresponds to more consistent results between different realizations.

In these panels, we show the time required to reach the stopping criterion for each system size. With the cold-start, at each system size we start from a randomly initialized network, while for the other protocols, we use the parameters for the system at half the size and use the corresponding transfer learning protocol. We remind the reader that for the transfer learning protocols, the time at size $L_v$ that we report is given by the accumulation of the time to obtain the state at size $L_v/2$ plus the time to reach the stopping criterion at size $L_v$, while the initial system size $L_v=4$ is obtained from a cold-start.   

For the Ising model, we observe that the $(1,2)-$tiling protocol reaches the stopping criterion in much longer times in the antiferromagnetic phase, even slower than a cold-start. The protocol is also not robust since we see a large variance in the error bars. Such performance is expected as the state in which the network is initialized does not favor an antiferromagnetic order as discussed in Sec.~\ref{sec:transfer_learning}. 
In general, for the Ising model, i.e. for Fig.\ref{fig:fig4} (a,c,e), we observe much better performance from the $(L,2)-$tiling, which also seems less sensitive to the phase of the system. We also observe that transfer learning protocols are more robust in the paramagnetic phase since the variance is very small.

For the Heisenberg model, we observe that the cold-start is always more efficient than the other transfer learning protocol, i.e. it reaches the stopping criterion the fastest. However efficiency does not necessarily imply effectiveness, and in fact, in this case the system is trapped in a local minimum prematurely. However, as we see later, this is not the case for the transfer learning protocols. We see that $(L,2)-$tiling protocol works efficiently except in the ferromagnetic phase (e.g for $\Delta=2$). This is expected as the state in which the network is initialized does not favor a ferromagnetic order when the magnetization is conserved, as discussed in Sec.~\ref{sec:transfer_learning}. Instead, the $(1,2)-$tiling and $(2,2)-$tiling protocols are more efficient in the ferromagnetic phase. Contrary to the Ising model, we observe that the transfer learning protocols are more robust in the ferromagnetic phase except for the $(L,2)-$tiling protocol.

\begin{figure}[t]
    \includegraphics[width=\columnwidth]{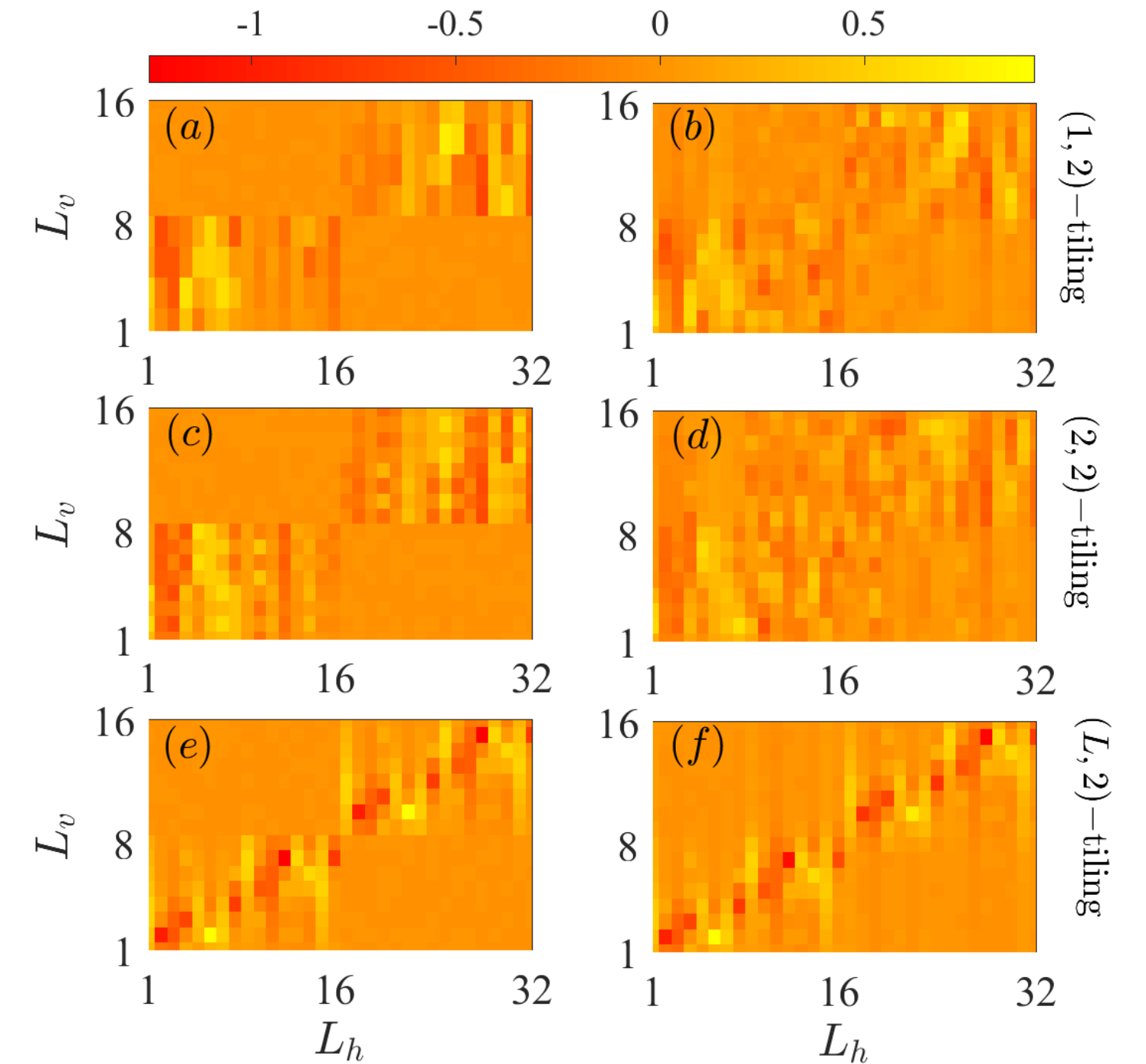}
    \caption{(Color online) Visualization of one realization of the weight matrices for one-dimensional ferromagnetic Ising model with 16 spins and $J_I$ = 2. The rows and columns of the matrix corresponds to the visible ($L_v$) and the hidden ($L_h$) nodes of the restricted Boltzmann machine, respectively. The left column panels (a), (c) and (e) show the initial weights $W^{(init)}$ after transferring from weights for an 8 spins chain with, respectively, the $(1,2)-$tiling, $(2,2)-$tiling and $(L,2)-$tiling protocols. The right column panels (b), (d) and (f) show the weights $W^{(final)}$ after fine-tuning until stopping criterion the initial weights $W^{(init)}$ in, respectively, panels (a), (c) and (e). The transfer distances $\mathcal{D}$ for the $(1,2)-$tiling, $(2,2)-$tiling and $(L,2)-$tiling protocol are $0.01145$, $0.01704$, and $0.00162$, respectively. 
    }
    \label{fig:fig5}
\end{figure}

\begin{figure}[t]
    \includegraphics[width=\columnwidth]{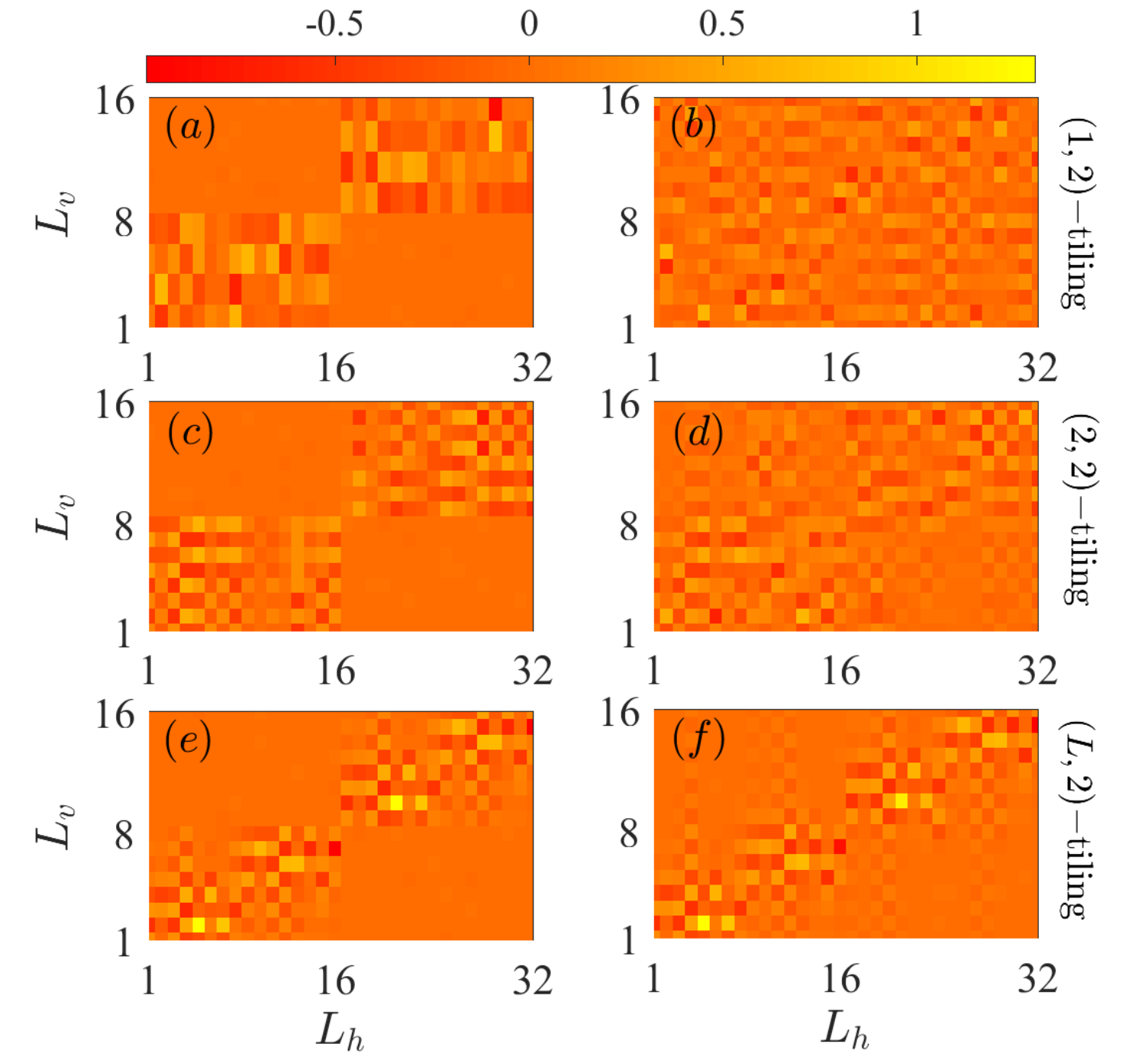}
    \caption{(Color online) Same as Fig.~\ref{fig:fig5} for the one-dimensional antiferromagnetic Ising model with 16 spins and $J_I$ = -2. 
    The transfer distances $\mathcal{D}$ for the $(1,2)-$tiling, $(2,2)-$tiling and $(L,2)-$tiling protocol are $0.05423$, $0.01858$, and $0.00224$, respectively.}
    \label{fig:fig6}
\end{figure}

We analyze the weights of the restricted Boltzmann machine $\bm{W}$ and, in particular, evaluate how they change when different transfer learning protocols are considered. In Fig.~\ref{fig:fig5} and Fig.~\ref{fig:fig6}, we consider one realization of the Ising model for ferromagnetic phase with $J_I=2$ and for antiferromagnetic phase with $J_I=-2$, respectively, until it reaches the stopping criterion. We depict the value of $W_{j,i}$, for a transfer from $L_v=8$ to $16$ spins where $\alpha = 2$, hence $j\in[1,16]$ and $i\in[1,32]$. The initial and final weights for the $(1,2)-$tiling protocol are shown respectively in Fig.~\ref{fig:fig5}(a,b) and Fig.~\ref{fig:fig6}(a,b), for the $(2,2)-$tiling protocol in Fig.~\ref{fig:fig5}(c,d) and Fig.~\ref{fig:fig6}(c,d), and for the $(L,2)-$tiling protocol in Fig.~\ref{fig:fig5}(e,f) and Fig.~\ref{fig:fig6}(e,f). From Fig.~\ref{fig:fig5} and \ref{fig:fig6}, we observe that the $(L,2)-$tiling protocol generically is a more adequate transfer protocol for Ising as it requires smaller changes in the weights to reach the stopping criterion. This is particularly striking when comparing Fig.\ref{fig:fig6}(a,b), from the $(1,2)-$tiling, with Fig.\ref{fig:fig6}(e,f), from $(L,2)-$tiling. The former requires a significant change off all the weights (resulting in $\mathcal{D}\approx 0.054$), while for the latter only some hidden nodes connections to some visible nodes are enhanced (corresponding to $\mathcal{D}\approx 0.002$, more than 25 times smaller than for the $(1,2)-$tiling). Qualitatively, we observe that the weight matrices in Fig.~\ref{fig:fig6} forms a checkerboard pattern which resemble the alternating spins in the antiferromagnetic phase. With the $(1,2)-$tiling protocol, the checkerboard pattern is destroyed, as seen in Fig.~\ref{fig:fig6}(a), and the network tries to restore this pattern, as seen in Fig.~\ref{fig:fig6}(b), which makes the transfer distance $\mathcal{D}$ bigger than with other protocols.  Quantitatively, the average transfer distances $\mathcal{D}$ over $20$ realizations for $(1,2)-$tiling, $(2,2)-$tiling, and $(L,2)-$tiling protocol from 64 to 128 spins in the antiferromagnetic phase with $J_I=-2.0$ are $6.5\times10^{-3}$,  $1.0\times10^{-3}$ and  $5.0\times10^{-5}$ , while in the ferromagnetic phase with $J_I=2.0$ they are $6.5\times10^{-4}$,  $9.6\times10^{-4}$ and  $5.0\times10^{-5}$, respectively \footnote{Here, and for the Heisenberg model discussed in the next paragraph, we have used the weights of a system with $64$ spins from the respective transfer learning protocol, as in Fig.\ref{fig:fig4}.}
These results show that the $(L,2)-$tiling protocol is the most adequate transfer learning protocol for the Ising model. 

For the Heisenberg model, over the 20 realizations, the average transfer distances $\mathcal{D}$ for $(1,2)-$tiling, $(2,2)-$tiling and $(L,2)-$tiling protocol for 128 spins in the antiferromagnetic phase  with $\Delta=-2$ are $4.9\times10^{-3}$, $4.5\times10^{-3}$ and $2.7\times10^{-4}$, while in the ferromagnetic phase  with $\Delta=2$ they are $1.6\times10^{-4}$, $2.0\times10^{-4}$ and $4.3\times10^{-4}$, respectively. These values confirm our previous results on the efficient evaluation of the Heisenberg model, showing that $(1,2)-$tiling protocol is the adequate transfer learning protocol for the ferromagnetic phase while $(L,2)-$tiling works best for the antiferromagnetic phase.  We do not plot the visualization of the weight matrices for the Heisenberg model as they show patterns that are similar to those of the Ising model.

\begin{figure}[t]
    \includegraphics[width=\columnwidth]{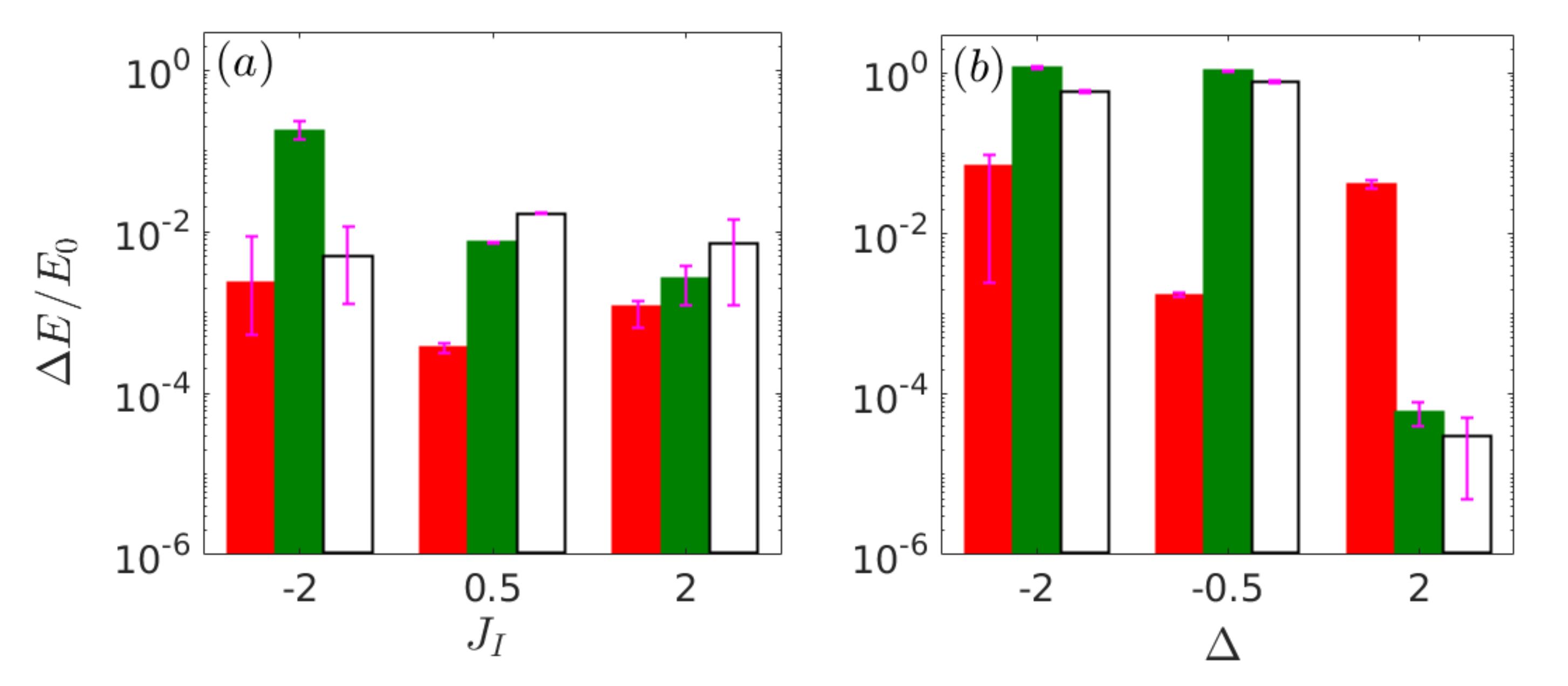}
    \caption{(Color online) The relative error of the energy ($\Delta E/E_0$) compared to the matrix product states simulations to measure the effectiveness of different transfer learning protocols. The evaluation is done on a one-dimensional system with 128 spins transferred from the fine-tuned weights of the most effective transfer learning protocol for 64 spins of the respective system and parameters. We evaluate the effectiveness at the iteration corresponding to the most efficient transfer learning protocol to reach the stopping criterion. Panels (a) and (b) correspond to the Ising and Heisenberg  models, respectively. In panel (a) we vary $J_I$ while in panel (b) we vary $\Delta$, and both take the values $\{-2, \pm 0.5, 2\}$. The red full bar, the black empty bar, and the green full bar show the relative error of the energy for the $(L,2)-$tiling, $(1,2)-$tiling, and $(2,2)-$tiling protocol, respectively. The error bars represent the interval of the 9th percentile and the 91st percentile.}
    \label{fig:fig7}
\end{figure}

While efficiency is an important attribute of a numerical method, effectiveness cannot be compromised, i.e. the representation of the ground state needs to be accurate. To evaluate the effectiveness, we consider the accuracy of the energy of the ground state and of the correlations generated. As a reference value, we take the calculations from very accurate matrix product states algorithm for the one-dimensional models. In Fig.~\ref{fig:fig7}(a), we measure the effectiveness for the ground state energy of the Ising model while in Fig.~\ref{fig:fig7}(b) we show the results from the Heisenberg model. In order to have a fair comparison between the different transfer learning protocols, we have considered the same base network which is the solution from the most effective protocol (i.e. the protocol with the lowest energy and correlation among others) for $L_v=64$, for the three transfer learning protocols $(1,2)-$tiling, $(2,2)-$tiling, and $(L,2)-$tiling protocol. We then use a fixed number of iterations which is given by the stopping criterion of the fastest transfer learning protocol to reach the stopping criterion. We observe trends that are similar to those found for the efficiency evaluation. For the Ising model, in general, the energy is more accurate with the $(L,2)-$tiling protocol, especially in the paramagnetic phase, while the $(1,2)-$tiling protocol, as expected, is particularly ineffective in the antiferromagnetic phase. Furthermore, by analyzing the error bars, we observe that the $(1,2)-$tiling protocol is less robust (has more variation from one realization to the next) in the antiferromagnetic phase whereas it is very robust in the paramagnetic phase.  Similarly, for the Heisenberg model, the energy is more accurate with $(L,2)-$tiling protocol except in the ferromagnetic phase. In our experiments, we notice that with $(1,2)-$tiling and $(2,2)-$tiling protocols, for the Heisenberg model with $\Delta = -0.5$ and $-2.0$, some realizations fail to converge because of the state initialized by the transfer that causes large gradients which result in the optimization algorithm to drive the system towards high energy states instead of minimizing the energy. These extreme cases are not taken into account for the calculation of the error bars presented in Fig.~\ref{fig:fig7}(b).

\begin{figure}[t]
    \includegraphics[width=\columnwidth]{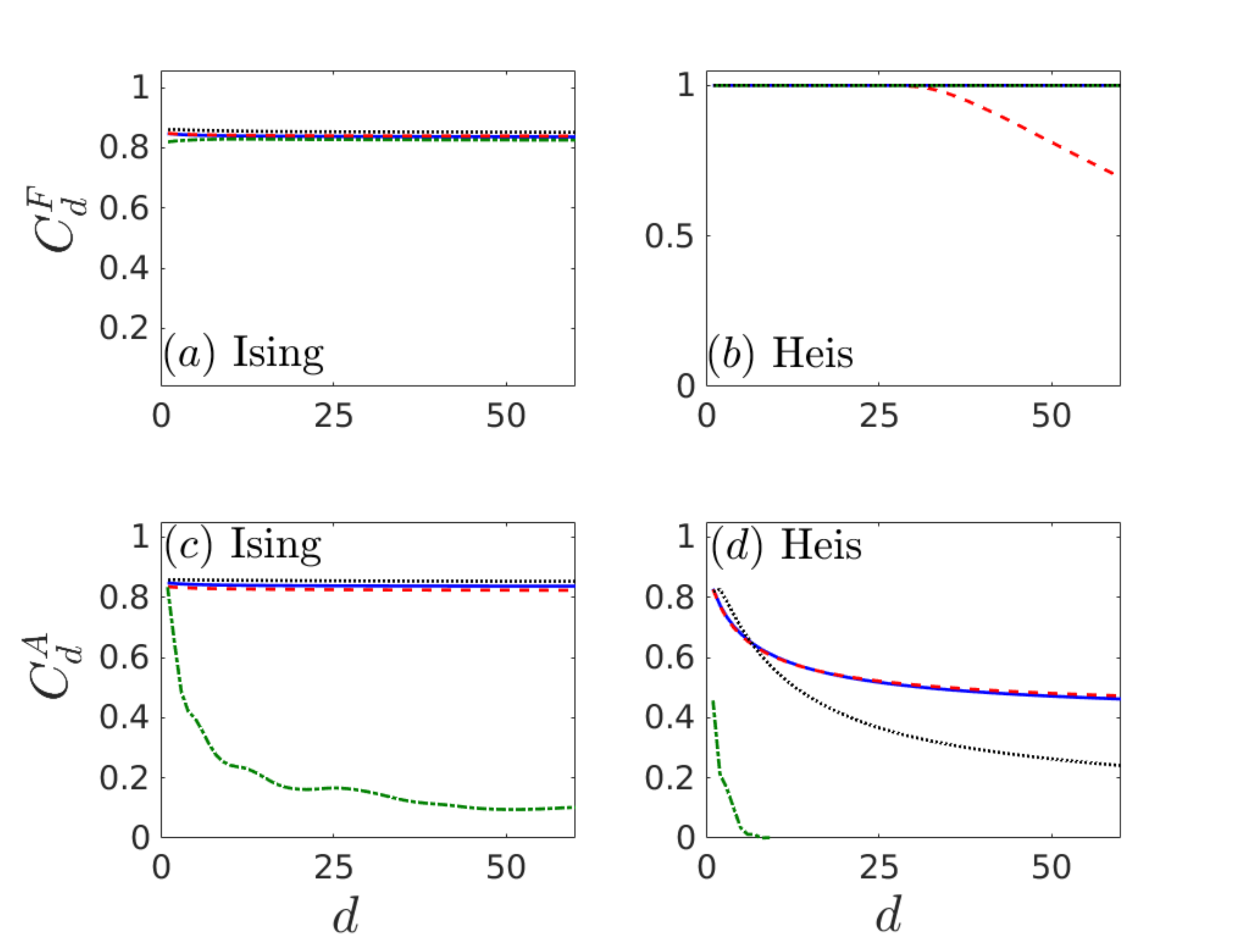}
    \caption{(Color online) Effectiveness in terms of the correlation between different transfer learning protocols. We evaluate the ferromagnetic correlator ($C^F_d$) and antiferromagnetic correlator ($C^A_d$) values for spins from distance $d=1$ to $d=60$ on one realization. This evaluation is done for a one-dimensional system with 128 spins in different phases transferred from the fine-tuned weights of the most effective transfer learning protocol for 64 spins of the respective system. We evaluate the effectiveness at the iteration of the most efficient transfer learning protocol to reach the stopping criterion. The left (a,c) and right (b,d) column panels show the correlator values as a function of the distance for the Ising and Heisenberg  model, respectively. The first row panels (a,b) show $C^F_d$ for ferromagnetic phases with $J_I=\Delta=2$, while the second row panels (c,d) show $C^A_d$ for antiferromagnetic phases with $J_I=\Delta=-2$. The blue solid line, red dashed line, the green dash-dotted line, and the black dotted  line correspond to the value from matrix product states simulations, $(L,2)-$tiling, $(1,2)-$tiling, and $(2,2)-$tiling protocols, respectively.}          
    \label{fig:fig8}
\end{figure}

\begin{figure}[t]
    \includegraphics[width=\columnwidth]{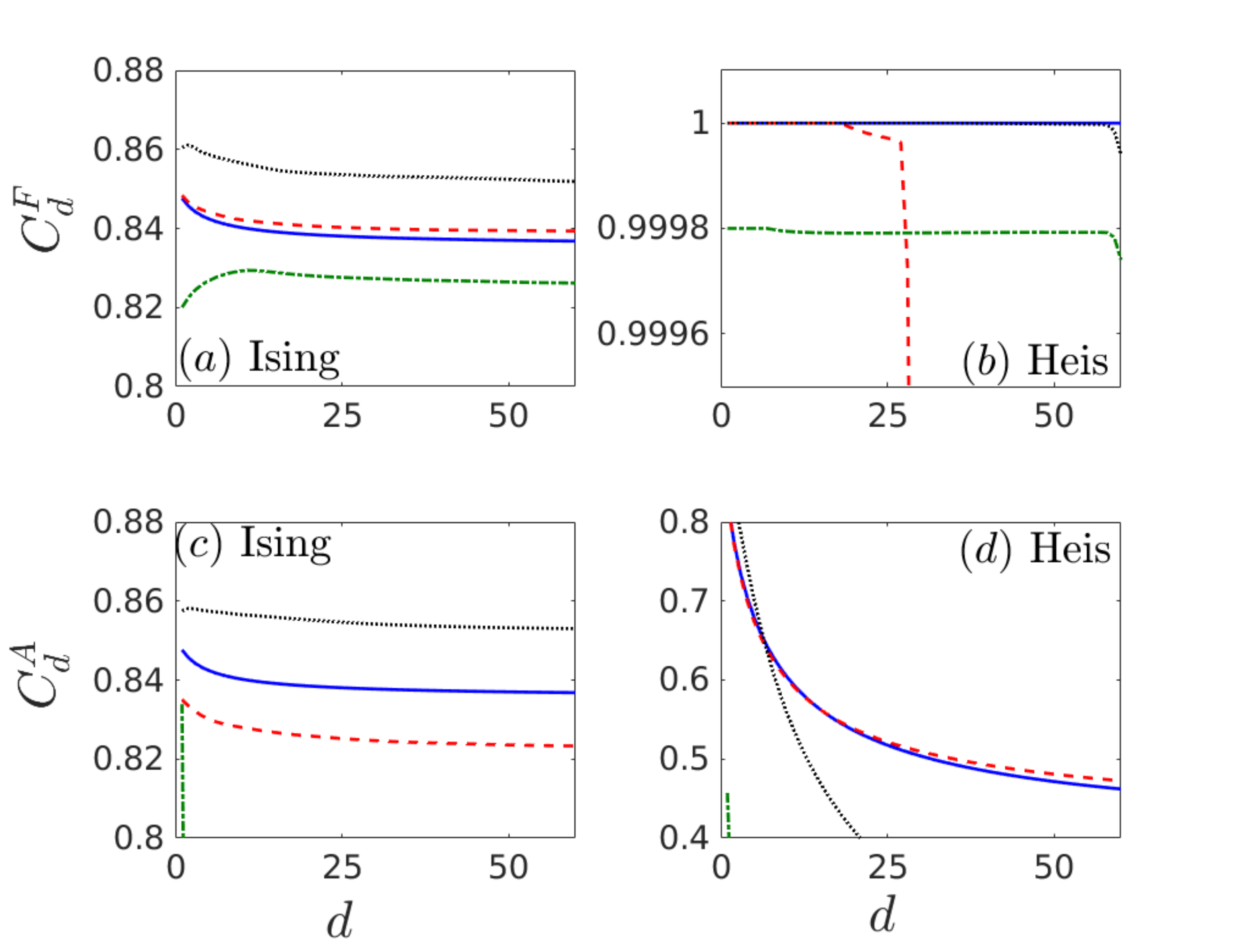}
    \caption{(Color online) This figure shows the zoomed-in version of Fig.~\ref{fig:fig8} to see the performance of each transfer learning protocol clearer. See the caption of Fig.~\ref{fig:fig8} for the complete description of the figure. }      
    \label{fig:fig9}
\end{figure}

In Fig.~\ref{fig:fig8}, we measure the effectiveness in terms of the correlation for one realization of the Ising and Heisenberg models in antiferromagnetic and ferromagnetic phases by plotting the correlators $C^A_d$ and $C^F_d$ as a function of the distance $d$ from 1 to 60. We observe in Fig.~\ref{fig:fig8}(b) that the correlator for the $(L,2)-$tiling protocol decays around $d\approx 30$. This ineffectiveness of $(L,2)-$tiling protocol is expected in this magnetization conserving case (see Sec.~\ref{sec:transfer_learning}) as we copied the configuration of a $L_v=64$ system which has a domain wall located around $d\simeq 32$. As shown earlier, we observe in Fig.~\ref{fig:fig8}(c,d) that the $(1,2)$-tiling protocol performs worst for both models in the antiferromagnetic phase. Figure~\ref{fig:fig9} shows the zoomed-in version of Fig.~\ref{fig:fig8} to compare which transfer learning protocol performs the best. In general, for both Ising and Heisenberg model, $(L,2)-$tiling protocol's line has the closest value to the matrix product states simulations except for the Heisenberg model in the ferromagnetic phase, for which the $(2,2)-$tiling protocol performs significantly better. As seen from the previous evaluations, we note that in this case also the error in energy is larger for the $(L,2)-$tiling protocol, and this transfer protocol is also the least efficient.     

\begin{figure}[t]
    \includegraphics[width=\columnwidth]{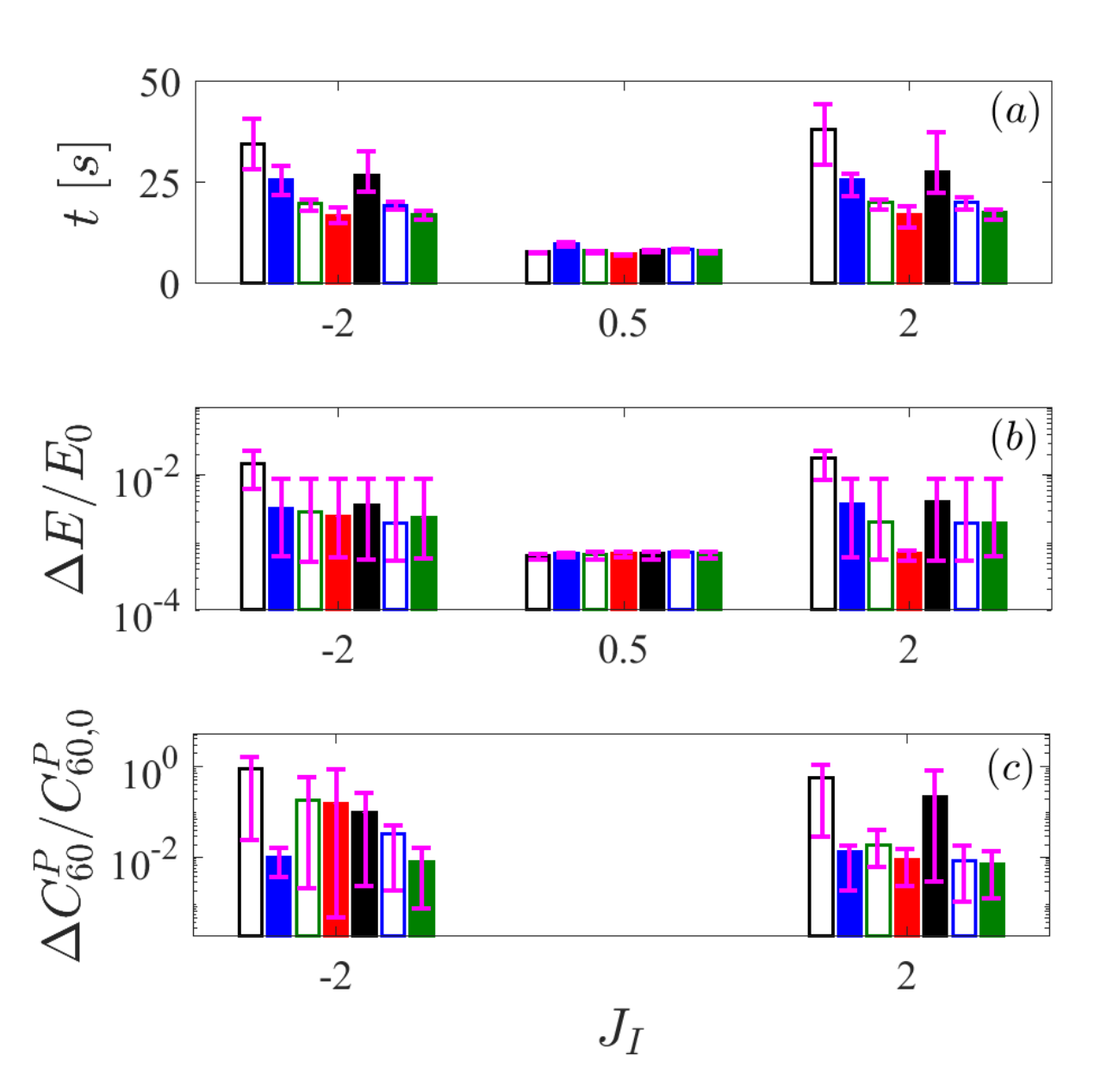}
    \caption{(Color online) The evaluation of different scenarios for $(L,p)-$tiling protocols, where $p = 2, 4, 8, 32$. The simulation is evaluated on one-dimensional Ising model at different Hamiltonian parameters ($J_I$) starting from 4 spins to get to 128 spins until the stopping criterion. (a) shows the efficiency of different scenarios in terms of time ($t[s]$). (b) shows the effectiveness in terms of the relative error for the energy ($\Delta E / E_0$) with the matrix product states simulation. (c) shows the effectiveness in terms of the relative error for correlations ($\Delta C^F_{60} / C^F_{60,0}$ for $J_I = 2$ and $\Delta C^A_{60} / C^A_{60,0}$ for $J_I = -2$) with the matrix product states simulation. The different scenarios are $4\rightarrow 128$ (black empty bar), $4\rightarrow 8\rightarrow 16\rightarrow 32\rightarrow 64\rightarrow 128$ (blue full bar), $4\rightarrow 16\rightarrow 128$ (green empty bar), $4\rightarrow 32\rightarrow 128$ (red full bar), $4\rightarrow 8\rightarrow 32\rightarrow 128$ (black full bar), $4\rightarrow 16\rightarrow 32\rightarrow 128$ (blue empty bar), and $4\rightarrow 16\rightarrow  64\rightarrow 128$ (green full bar).  The error bars represent the interval of the 9th percentile and the 91st percentile. 
    }
    \label{fig:fig10}
\end{figure}

Up to now, we have considered scenarios in which the size of the system was doubled all the time. However, it is also possible to use these transfer learning strategies to increase the size of the system by other increments as, for instance, a factor of $8$. In Fig.~\ref{fig:fig10}, we consider several scenarios to reach $128$ spins in the one-dimensional Ising model starting from $4$ spins, i.e. transferring from $4\rightarrow 128$, $(4,32)-$tiling (black empty bar),  $4\rightarrow 8\rightarrow 16\rightarrow 32\rightarrow 64\rightarrow 128$ (blue full bar), $4\rightarrow 16\rightarrow 128$ (green empty bar), $4\rightarrow 32\rightarrow 128$ (red full bar), $4\rightarrow 8\rightarrow 32\rightarrow 128$ (black full bar), $4\rightarrow 16\rightarrow 32\rightarrow 128$ (blue empty bar) and $4\rightarrow 16\rightarrow  64\rightarrow 128$ (green full bar). In all these steps, we have used the $(L,p)-$tiling protocol with $p=2,\;4,\;8$ or $32$, and we run the algorithm until the stopping criterion is reached. We observe that a large increase in the $4\rightarrow 128$ scenario performs the worst in ferromagnetic and antiferromagnetic phases. This is because the small systems do not have enough knowledge of the correlations to readily approximate the large system. Despite this, we note that the $4\rightarrow 128$ scenario is still better than from a cold-start. As a reference, the average time for cold-start in the antiferromagnetic phase is $291.706$ seconds which is about $8\times$ slower than the $4\rightarrow 128$ scenario. Furthermore, the relative error of the energy of the $4\rightarrow 128$ scenario is $30\%$ better than the cold-start. 

We observe that there is a trade-off between efficiency and effectiveness in the choice of the transfer learning protocols. Doubling each time the size with five increments to reach $L_v=128$ is an effective but not an efficient scenario. As an alternative, we can choose the scenario $4\rightarrow 16\rightarrow  64\rightarrow 128$ that is competitively effective but more efficient. If we aim for an efficient scenario, then the scenario $4\rightarrow 32\rightarrow 128$ is the most efficient but not as effective. However, we observe that this depends on the phase. Scenario $4\rightarrow 32\rightarrow 128$ is the most efficient and effective transfer in the ferromagnetic phase but not in the antiferromagnetic one. In the paramagnetic phase, all scenarios work similarly well. In terms of the size of the increment, we observe that a big increase in an early transfer is preferred to one in a later transfer. For instance, the scenario $4\rightarrow 32\rightarrow 128$ is more efficient and effective than $4\rightarrow 16\rightarrow 128$.

\begin{figure}[t]
    \includegraphics[width=\columnwidth]{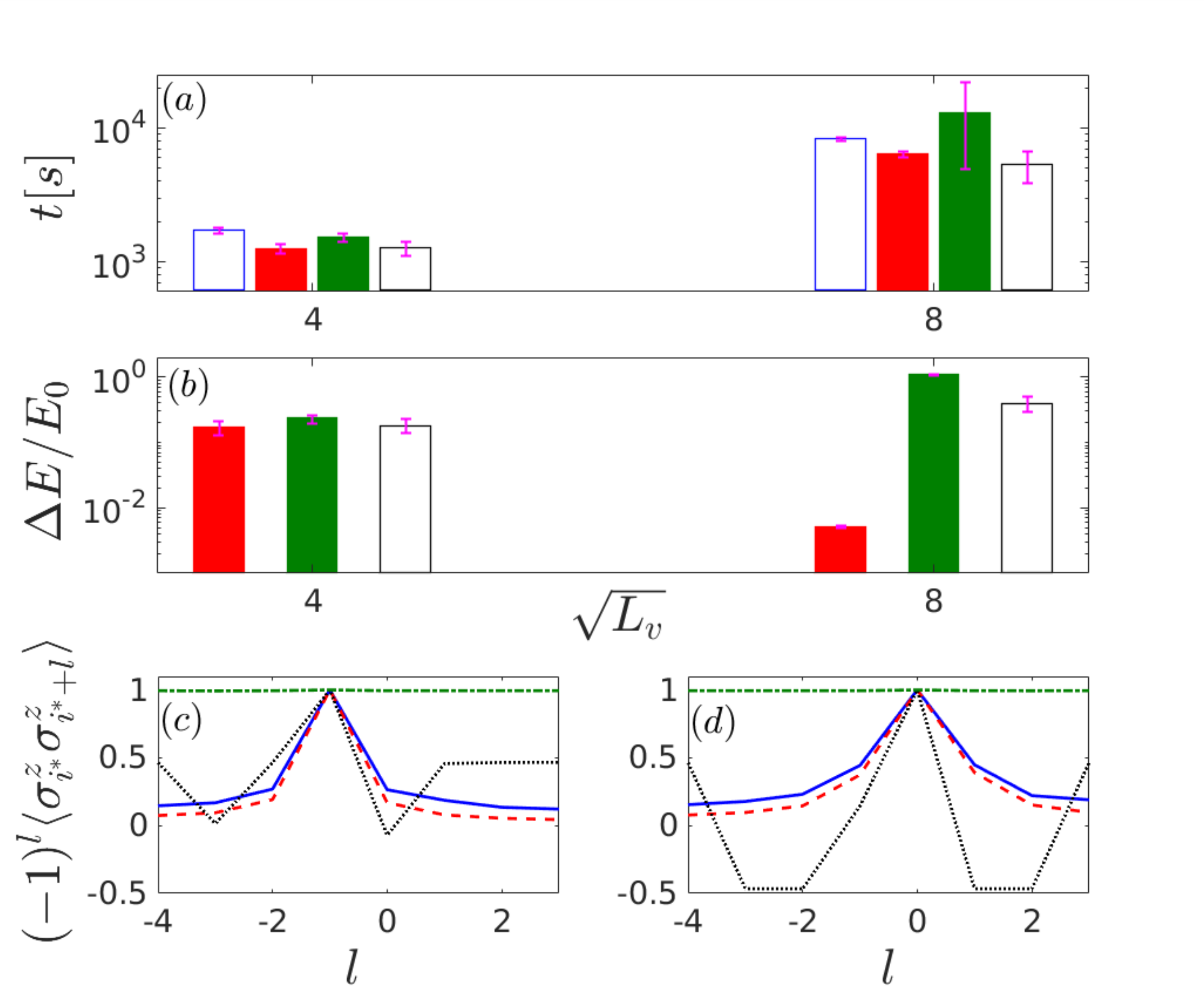}
    \caption{(Color online) The evaluation of the efficiency and the effectiveness of different transfer learning protocols in a two-dimensional system. The evaluation is done on the Heisenberg model with $J_{XXZ}=1$ and $\Delta=-1$. Panel (a) shows the efficiency of different transfer learning protocols. The panel shows the simulation time in seconds ($t[s]$) until the stopping criterion as a function of the length $\sqrt{L_v} = 4$ or $8$. The blue empty bars are for cold-start, red full bars are for $(L,2)-$tiling protocol, green full bars for $(1,2)-$tiling protocol, and black empty bars for $(2,2)-$tiling protocol.  
    Panel (b) shows the effectiveness of different transfer learning protocols in terms of the ground state energy. The panel shows the mean relative error of different transfer learning protocols for systems with $4\times4$ and $8\times8$ particles. To evaluate the effectiveness, we use results from quantum Monte Carlo simulations. Note that the error in energy for the $4\times4$ system has been multiplied by 200 to make it clearer in the figure. Panels (c,d) show the effectiveness of different transfer learning protocols in terms of the correlation for a  $8\times8$ system. The panel shows antiferromagnetic correlator $(-1)^l\langle\sigma^z_{i^*}\sigma^z_{i^*+l}\rangle$ where $i^*$ is the site $(4,5)$ and $l$ is the distance from this site either on the diagonal, panel (c), or on the same row, panel (d). The solid blue line corresponds to quantum Monte Carlo simulations,  the red dashed line corresponds to the $(L,2)-$tiling protocol, the green dash-dotted line corresponds to $(1,2)-$tiling protocol and the black dotted line corresponds to $(2,2)-$tiling protocol. 
   }
    \label{fig:fig11}
\end{figure}

Our study is not confined to one-dimensional systems, and the transfer learning can also be adopted, for instance, in two-dimensional systems. 

We thus consider the Heisenberg model with open boundary conditions, $\Delta=-1$ and $J_{XXZ}=1$ and consider sizes $2\times 2$, $4\times 4$ and $8\times 8$. As mentioned earlier, this is equivalent to the isotropic antiferromagnetic Heisenberg model. In Fig.~\ref{fig:fig11}, we show the time needed for the calculations for the different transfer learning protocols, panel (a), the error on the energy in the different phases of systems with $4\times 4$ and $8\times 8$ spins, panel (b), and the antiferromagnetic correlator $(-1^l)\langle \sigma^z_{i^*}\sigma^z_{i^*+l} \rangle$ as a function of the distance between spins $l$ for a system with $8\times 8$ spins, panels (c,d). In particular, we have chosen the position $i^*$ to be on the $4-$th row and $5-$th column of the system (in the center) and we have taken the other spins to be either on the diagonal, panel (c), or on the same row, panel (d).              
We use the values from quantum Monte Carlo simulations~\cite{sandvik1997finite} as references. Similarly to the one-dimensional case, we see that the $(L,2)-$tiling protocol is the most efficient and effective protocol followed by $(2,2)-$tiling and $(1,2)-$tiling protocols in the antiferromagnetic phase.  We also observe that all the transfer learning protocols are quite robust, since the error bars are relatively small, except for the time required to reach the stopping criterion for the $(1,2)-$tiling protocol.

\section{Conclusions} \label{sec:conclusions}

We have proposed the idea of transfer learning for the scalability of neural-network quantum states. Among a multitude of candidates in the design space of  transfer learning protocols for the problem at hand, we devised and presented a selected few that seem to be amenable to physical interpretation. We comparatively evaluated the performance of the different transfer learning protocols with respect to the efficiency (speed) and the effectiveness (accuracy) of neural-network quantum states in different phases, for different models in both one and two dimensions. 

We have considered two prototypical models, the transverse field Ising and the Heisenberg XXZ model. The two models have similarities in the phases they manifest but also have an important difference. Namely, the space of configurations of the Heisenberg XXZ model is reduced because it conserves the total magnetization and we constrained it to a fixed total magnetization sector, while all possible values of the magnetization are accessible starting from any state for the Ising model. Because of this difference, each transfer learning protocol performs differently in the two models.

The transfer learning protocol that we call $(L,2)-$tiling is the best performing protocol for the transverse field Ising one-dimensional model and is generally competitive for the Heisenberg XXZ one-dimensional and two-dimensional model. The protocol is less efficient and effective than other protocols that we call  $(1,2)-$tiling and  $(2,2)-$tiling  for the ferromagnetic Heisenberg XXZ model due to the zero magnetization constraint imposed on the system. We also investigated scenarios other than doubling the size of the system. In these cases, we observed that there is a trade-off between efficiency and effectiveness between different jumping scenarios. We found that, in general, larger increases of system size at the beginning of the transfer learning and smaller increases later are preferable. These observations suggest the need for further studies to precisely characterize the physical principles that determine the success of a given transfer learning protocol depending on the system and state considered.  

In summary, our empirical results demonstrate that transfer learning for the scalability of neural-network quantum states can bring two advantages: (i) efficiency - it allows to reach a good approximation of the ground state in shorter time compared to a cold-start, and (ii) effectiveness - it reduces the chances that the optimization procedure remains trapped in a local minima far from the ground state. 

For the transfer learning to be productive, a tiling protocol that is adapted to the model and the phase that it adopts must be chosen. If the patterns that are present in the phase are not preserved in the transfer, the method becomes inefficient and ineffective, as is the case for the $(1,2)-$tiling protocol in the antiferromagnetic phase. On the contrary, transfer learning protocol that preserves such patterns generally provide good scalability.

Our exploration of possible tiling protocol is of course not exhaustive and other protocols could provide even better scalability. We are currently investigating other opportunities to leverage transfer learning for neural-network quantum states. We are investigating transfer learning between  neural-network quantum states for systems of equal size but different parameter regimes of the same Hamiltonian as well as between  neural-network quantum states for systems with different Hamiltonians. We are also considering transfer on  restricted Boltzmann machines with complex-valued weights and with symmetries such as translational invariance.

\section*{Acknowledgement} 
We acknowledge C. Guo and Supremacy Future Technologies for support on the matrix product states simulations. This work is partially funded by the National University of Singapore, the French Ministry of European and Foreign Affairs and the French Ministry of Higher Education, Research and Innovation under the Merlion programme as Merlion Project ``Deep Quantum''.
Some of the experiments reported in this article are performed on the infrastructure of Singapore National Supercomputing Centre \cite{nscc} and are funded under project ``Computing the Deep Quantum''.

\bibliography{apssamp}

\begin{thebibliography}{29}%
\makeatletter
\providecommand \@ifxundefined [1]{%
 \@ifx{#1\undefined}
}%
\providecommand \@ifnum [1]{%
 \ifnum #1\expandafter \@firstoftwo
 \else \expandafter \@secondoftwo
 \fi
}%
\providecommand \@ifx [1]{%
 \ifx #1\expandafter \@firstoftwo
 \else \expandafter \@secondoftwo
 \fi
}%
\providecommand \natexlab [1]{#1}%
\providecommand \enquote  [1]{``#1''}%
\providecommand \bibnamefont  [1]{#1}%
\providecommand \bibfnamefont [1]{#1}%
\providecommand \citenamefont [1]{#1}%
\providecommand \href@noop [0]{\@secondoftwo}%
\providecommand \href [0]{\begingroup \@sanitize@url \@href}%
\providecommand \@href[1]{\@@startlink{#1}\@@href}%
\providecommand \@@href[1]{\endgroup#1\@@endlink}%
\providecommand \@sanitize@url [0]{\catcode `\\12\catcode `\$12\catcode
  `\&12\catcode `\#12\catcode `\^12\catcode `\_12\catcode `\%12\relax}%
\providecommand \@@startlink[1]{}%
\providecommand \@@endlink[0]{}%
\providecommand \url  [0]{\begingroup\@sanitize@url \@url }%
\providecommand \@url [1]{\endgroup\@href {#1}{\urlprefix }}%
\providecommand \urlprefix  [0]{URL }%
\providecommand \Eprint [0]{\href }%
\providecommand \doibase [0]{http://dx.doi.org/}%
\providecommand \selectlanguage [0]{\@gobble}%
\providecommand \bibinfo  [0]{\@secondoftwo}%
\providecommand \bibfield  [0]{\@secondoftwo}%
\providecommand \translation [1]{[#1]}%
\providecommand \BibitemOpen [0]{}%
\providecommand \bibitemStop [0]{}%
\providecommand \bibitemNoStop [0]{.\EOS\space}%
\providecommand \EOS [0]{\spacefactor3000\relax}%
\providecommand \BibitemShut  [1]{\csname bibitem#1\endcsname}%
\let\auto@bib@innerbib\@empty
%</preamble>
\bibitem [{\citenamefont {Gubernatis}\ \emph {et~al.}(2016)\citenamefont
  {Gubernatis}, \citenamefont {Kawashima},\ and\ \citenamefont
  {Werner}}]{gubernatis2016}%
  \BibitemOpen
  \bibfield  {author} {\bibinfo {author} {\bibfnamefont {J.}~\bibnamefont
  {Gubernatis}}, \bibinfo {author} {\bibfnamefont {N.}~\bibnamefont
  {Kawashima}}, \ and\ \bibinfo {author} {\bibfnamefont {P.}~\bibnamefont
  {Werner}},\ }\href {\doibase 10.1017/CBO9780511902581} {\emph {\bibinfo
  {title} {Quantum Monte Carlo Methods: Algorithms for Lattice Models}}}\
  (\bibinfo  {publisher} {Cambridge University Press},\ \bibinfo {year}
  {2016})\BibitemShut {NoStop}%
\bibitem [{\citenamefont {Sandvik}(1997)}]{sandvik1997finite}%
  \BibitemOpen
  \bibfield  {author} {\bibinfo {author} {\bibfnamefont {A.~W.}\ \bibnamefont
  {Sandvik}},\ }\href@noop {} {\bibfield  {journal} {\bibinfo  {journal}
  {Physical Review B}\ }\textbf {\bibinfo {volume} {56}},\ \bibinfo {pages}
  {11678} (\bibinfo {year} {1997})}\BibitemShut {NoStop}%
\bibitem [{\citenamefont {Or{\'u}s}(2014)}]{orus2014practical}%
  \BibitemOpen
  \bibfield  {author} {\bibinfo {author} {\bibfnamefont {R.}~\bibnamefont
  {Or{\'u}s}},\ }\href@noop {} {\bibfield  {journal} {\bibinfo  {journal} {Ann.
  Phys.}\ }\textbf {\bibinfo {volume} {349}},\ \bibinfo {pages} {117} (\bibinfo
  {year} {2014})}\BibitemShut {NoStop}%
\bibitem [{\citenamefont {Verstraete}\ \emph {et~al.}(2008)\citenamefont
  {Verstraete}, \citenamefont {Cirac},\ and\ \citenamefont
  {Murg}}]{verstraete2004renormalization}%
  \BibitemOpen
  \bibfield  {author} {\bibinfo {author} {\bibfnamefont {F.}~\bibnamefont
  {Verstraete}}, \bibinfo {author} {\bibfnamefont {J.~I.}\ \bibnamefont
  {Cirac}}, \ and\ \bibinfo {author} {\bibfnamefont {V.}~\bibnamefont {Murg}},\
  }\href@noop {} {\bibfield  {journal} {\bibinfo  {journal} {Adv. Phys.}\
  }\textbf {\bibinfo {volume} {57}},\ \bibinfo {pages} {143} (\bibinfo {year}
  {2008})}\BibitemShut {NoStop}%
\bibitem [{\citenamefont {Schollw{\"o}ck}(2005)}]{schollwock2005density}%
  \BibitemOpen
  \bibfield  {author} {\bibinfo {author} {\bibfnamefont {U.}~\bibnamefont
  {Schollw{\"o}ck}},\ }\href@noop {} {\bibfield  {journal} {\bibinfo  {journal}
  {Rev. Mod. Phys.}\ }\textbf {\bibinfo {volume} {77}},\ \bibinfo {pages} {259}
  (\bibinfo {year} {2005})}\BibitemShut {NoStop}%
\bibitem [{\citenamefont {Schollw\"ock}(2011)}]{schollwock2011}%
  \BibitemOpen
  \bibfield  {author} {\bibinfo {author} {\bibfnamefont {U.}~\bibnamefont
  {Schollw\"ock}},\ }\href {\doibase https://doi.org/10.1016/j.aop.2010.09.012}
  {\bibfield  {journal} {\bibinfo  {journal} {Ann. Phys.}\ }\textbf {\bibinfo
  {volume} {326}},\ \bibinfo {pages} {96 } (\bibinfo {year}
  {2011})}\BibitemShut {NoStop}%
\bibitem [{\citenamefont {White}(1992)}]{white1992density}%
  \BibitemOpen
  \bibfield  {author} {\bibinfo {author} {\bibfnamefont {S.~R.}\ \bibnamefont
  {White}},\ }\href@noop {} {\bibfield  {journal} {\bibinfo  {journal} {Phys.
  Rev. Lett.}\ }\textbf {\bibinfo {volume} {69}},\ \bibinfo {pages} {2863}
  (\bibinfo {year} {1992})}\BibitemShut {NoStop}%
\bibitem [{\citenamefont {Georges}\ \emph {et~al.}(1996)\citenamefont
  {Georges}, \citenamefont {Kotliar}, \citenamefont {Krauth},\ and\
  \citenamefont {Rozenberg}}]{georges1996}%
  \BibitemOpen
  \bibfield  {author} {\bibinfo {author} {\bibfnamefont {A.}~\bibnamefont
  {Georges}}, \bibinfo {author} {\bibfnamefont {G.}~\bibnamefont {Kotliar}},
  \bibinfo {author} {\bibfnamefont {W.}~\bibnamefont {Krauth}}, \ and\ \bibinfo
  {author} {\bibfnamefont {M.~J.}\ \bibnamefont {Rozenberg}},\ }\href@noop {}
  {\bibfield  {journal} {\bibinfo  {journal} {Rev. Mod. Phys.}\ }\textbf
  {\bibinfo {volume} {68}},\ \bibinfo {pages} {13} (\bibinfo {year}
  {1996})}\BibitemShut {NoStop}%
\bibitem [{\citenamefont {Metzner}\ and\ \citenamefont
  {Vollhardt}(1989)}]{Vollhardt1989}%
  \BibitemOpen
  \bibfield  {author} {\bibinfo {author} {\bibfnamefont {W.}~\bibnamefont
  {Metzner}}\ and\ \bibinfo {author} {\bibfnamefont {D.}~\bibnamefont
  {Vollhardt}},\ }\href {\doibase 10.1103/PhysRevLett.62.324} {\bibfield
  {journal} {\bibinfo  {journal} {Phys. Rev. Lett.}\ }\textbf {\bibinfo
  {volume} {62}},\ \bibinfo {pages} {324} (\bibinfo {year} {1989})}\BibitemShut
  {NoStop}%
\bibitem [{\citenamefont {Georges}\ and\ \citenamefont
  {Kotliar}(1992)}]{GeorgesKotliar1992}%
  \BibitemOpen
  \bibfield  {author} {\bibinfo {author} {\bibfnamefont {A.}~\bibnamefont
  {Georges}}\ and\ \bibinfo {author} {\bibfnamefont {G.}~\bibnamefont
  {Kotliar}},\ }\href {\doibase 10.1103/PhysRevB.45.6479} {\bibfield  {journal}
  {\bibinfo  {journal} {Phys. Rev. B}\ }\textbf {\bibinfo {volume} {45}},\
  \bibinfo {pages} {6479} (\bibinfo {year} {1992})}\BibitemShut {NoStop}%
\bibitem [{\citenamefont {Carleo}\ and\ \citenamefont
  {Troyer}(2017)}]{carleo2017solving}%
  \BibitemOpen
  \bibfield  {author} {\bibinfo {author} {\bibfnamefont {G.}~\bibnamefont
  {Carleo}}\ and\ \bibinfo {author} {\bibfnamefont {M.}~\bibnamefont
  {Troyer}},\ }\href@noop {} {\bibfield  {journal} {\bibinfo  {journal}
  {Science}\ }\textbf {\bibinfo {volume} {355}},\ \bibinfo {pages} {602}
  (\bibinfo {year} {2017})}\BibitemShut {NoStop}%
\bibitem [{\citenamefont {Choo}\ \emph {et~al.}(2018)\citenamefont {Choo},
  \citenamefont {Carleo}, \citenamefont {Regnault},\ and\ \citenamefont
  {Neupert}}]{choo2018symmetries}%
  \BibitemOpen
  \bibfield  {author} {\bibinfo {author} {\bibfnamefont {K.}~\bibnamefont
  {Choo}}, \bibinfo {author} {\bibfnamefont {G.}~\bibnamefont {Carleo}},
  \bibinfo {author} {\bibfnamefont {N.}~\bibnamefont {Regnault}}, \ and\
  \bibinfo {author} {\bibfnamefont {T.}~\bibnamefont {Neupert}},\ }\href@noop
  {} {\bibfield  {journal} {\bibinfo  {journal} {Phys. Rev. Lett.}\ }\textbf
  {\bibinfo {volume} {121}},\ \bibinfo {pages} {167204} (\bibinfo {year}
  {2018})}\BibitemShut {NoStop}%
\bibitem [{\citenamefont {Czischek}\ \emph {et~al.}(2018)\citenamefont
  {Czischek}, \citenamefont {G{\"a}rttner},\ and\ \citenamefont
  {Gasenzer}}]{czischek2018quenches}%
  \BibitemOpen
  \bibfield  {author} {\bibinfo {author} {\bibfnamefont {S.}~\bibnamefont
  {Czischek}}, \bibinfo {author} {\bibfnamefont {M.}~\bibnamefont
  {G{\"a}rttner}}, \ and\ \bibinfo {author} {\bibfnamefont {T.}~\bibnamefont
  {Gasenzer}},\ }\href@noop {} {\bibfield  {journal} {\bibinfo  {journal}
  {Phys. Rev. B}\ }\textbf {\bibinfo {volume} {98}},\ \bibinfo {pages} {024311}
  (\bibinfo {year} {2018})}\BibitemShut {NoStop}%
\bibitem [{\citenamefont {Deng}\ \emph {et~al.}(2017)\citenamefont {Deng},
  \citenamefont {Li},\ and\ \citenamefont {Sarma}}]{deng2017quantum}%
  \BibitemOpen
  \bibfield  {author} {\bibinfo {author} {\bibfnamefont {D.-L.}\ \bibnamefont
  {Deng}}, \bibinfo {author} {\bibfnamefont {X.}~\bibnamefont {Li}}, \ and\
  \bibinfo {author} {\bibfnamefont {S.~D.}\ \bibnamefont {Sarma}},\ }\href@noop
  {} {\bibfield  {journal} {\bibinfo  {journal} {Phys. Rev. X}\ }\textbf
  {\bibinfo {volume} {7}},\ \bibinfo {pages} {021021} (\bibinfo {year}
  {2017})}\BibitemShut {NoStop}%
\bibitem [{\citenamefont {Melko}\ \emph {et~al.}(2019)\citenamefont {Melko},
  \citenamefont {Carleo}, \citenamefont {Carrasquilla},\ and\ \citenamefont
  {Cirac}}]{melko2019restricted}%
  \BibitemOpen
  \bibfield  {author} {\bibinfo {author} {\bibfnamefont {R.}~\bibnamefont
  {Melko}}, \bibinfo {author} {\bibfnamefont {G.}~\bibnamefont {Carleo}},
  \bibinfo {author} {\bibfnamefont {J.}~\bibnamefont {Carrasquilla}}, \ and\
  \bibinfo {author} {\bibfnamefont {J.~I.}\ \bibnamefont {Cirac}},\ }\href@noop
  {} {\bibfield  {journal} {\bibinfo  {journal} {Nature Physics}\ } (\bibinfo
  {year} {2019})}\BibitemShut {NoStop}%
\bibitem [{\citenamefont {Das~Sarma}\ \emph {et~al.}(2019)\citenamefont
  {Das~Sarma}, \citenamefont {Deng},\ and\ \citenamefont
  {Duan}}]{sarma2019review}%
  \BibitemOpen
  \bibfield  {author} {\bibinfo {author} {\bibfnamefont {S.}~\bibnamefont
  {Das~Sarma}}, \bibinfo {author} {\bibfnamefont {D.-L.}\ \bibnamefont {Deng}},
  \ and\ \bibinfo {author} {\bibfnamefont {L.-M.}\ \bibnamefont {Duan}},\
  }\href@noop {} {\bibfield  {journal} {\bibinfo  {journal} {Physics Today}\
  }\textbf {\bibinfo {volume} {72}},\ \bibinfo {pages} {48} (\bibinfo {year}
  {2019})}\BibitemShut {NoStop}%
\bibitem [{\citenamefont {Carleo}\ \emph
  {et~al.}(2019{\natexlab{a}})\citenamefont {Carleo}, \citenamefont {Cirac},
  \citenamefont {Cranmer}, \citenamefont {Daudet}, \citenamefont {Schuld},
  \citenamefont {Tishby}, \citenamefont {Vogt-Maranto},\ and\ \citenamefont
  {Zdeborov\'a}}]{Carleo2019review}%
  \BibitemOpen
  \bibfield  {author} {\bibinfo {author} {\bibfnamefont {G.}~\bibnamefont
  {Carleo}}, \bibinfo {author} {\bibfnamefont {J.~I.}\ \bibnamefont {Cirac}},
  \bibinfo {author} {\bibfnamefont {K.}~\bibnamefont {Cranmer}}, \bibinfo
  {author} {\bibfnamefont {L.}~\bibnamefont {Daudet}}, \bibinfo {author}
  {\bibfnamefont {M.}~\bibnamefont {Schuld}}, \bibinfo {author} {\bibfnamefont
  {N.}~\bibnamefont {Tishby}}, \bibinfo {author} {\bibfnamefont
  {L.}~\bibnamefont {Vogt-Maranto}}, \ and\ \bibinfo {author} {\bibfnamefont
  {L.}~\bibnamefont {Zdeborov\'a}},\ }\href@noop {} {\bibfield  {journal}
  {\bibinfo  {journal} {arxiv:1903.10563}\ } (\bibinfo {year}
  {2019}{\natexlab{a}})}\BibitemShut {NoStop}%
\bibitem [{\citenamefont {Choo}\ \emph {et~al.}(2019)\citenamefont {Choo},
  \citenamefont {Neupert},\ and\ \citenamefont {Carleo}}]{ChooCarleo2019}%
  \BibitemOpen
  \bibfield  {author} {\bibinfo {author} {\bibfnamefont {K.}~\bibnamefont
  {Choo}}, \bibinfo {author} {\bibfnamefont {T.}~\bibnamefont {Neupert}}, \
  and\ \bibinfo {author} {\bibfnamefont {G.}~\bibnamefont {Carleo}},\
  }\href@noop {} {\bibfield  {journal} {\bibinfo  {journal} {arxiv:1903.06713}\
  } (\bibinfo {year} {2019})}\BibitemShut {NoStop}%
\bibitem [{\citenamefont {Carleo}\ \emph
  {et~al.}(2019{\natexlab{b}})\citenamefont {Carleo}, \citenamefont {Choo},
  \citenamefont {Hofmann}, \citenamefont {Smith}, \citenamefont {Westerhout},
  \citenamefont {Alet}, \citenamefont {Davis}, \citenamefont {Efthymiou},
  \citenamefont {Glasser}, \citenamefont {Lin} \emph
  {et~al.}}]{carleo2019netket}%
  \BibitemOpen
  \bibfield  {author} {\bibinfo {author} {\bibfnamefont {G.}~\bibnamefont
  {Carleo}}, \bibinfo {author} {\bibfnamefont {K.}~\bibnamefont {Choo}},
  \bibinfo {author} {\bibfnamefont {D.}~\bibnamefont {Hofmann}}, \bibinfo
  {author} {\bibfnamefont {J.~E.}\ \bibnamefont {Smith}}, \bibinfo {author}
  {\bibfnamefont {T.}~\bibnamefont {Westerhout}}, \bibinfo {author}
  {\bibfnamefont {F.}~\bibnamefont {Alet}}, \bibinfo {author} {\bibfnamefont
  {E.~J.}\ \bibnamefont {Davis}}, \bibinfo {author} {\bibfnamefont
  {S.}~\bibnamefont {Efthymiou}}, \bibinfo {author} {\bibfnamefont
  {I.}~\bibnamefont {Glasser}}, \bibinfo {author} {\bibfnamefont {S.-H.}\
  \bibnamefont {Lin}},  \emph {et~al.},\ }\href@noop {} {\bibfield  {journal}
  {\bibinfo  {journal} {arXiv:1904.00031}\ } (\bibinfo {year}
  {2019}{\natexlab{b}})}\BibitemShut {NoStop}%
\bibitem [{\citenamefont {Dietterich}\ \emph {et~al.}(1997)\citenamefont
  {Dietterich}, \citenamefont {Pratt},\ and\ \citenamefont
  {Thrun}}]{dietterich1997special}%
  \BibitemOpen
  \bibfield  {author} {\bibinfo {author} {\bibfnamefont {T.~G.}\ \bibnamefont
  {Dietterich}}, \bibinfo {author} {\bibfnamefont {L.}~\bibnamefont {Pratt}}, \
  and\ \bibinfo {author} {\bibfnamefont {S.}~\bibnamefont {Thrun}},\
  }\href@noop {} {\bibfield  {journal} {\bibinfo  {journal} {Mach. Learn.}\
  }\textbf {\bibinfo {volume} {28}} (\bibinfo {year} {1997})}\BibitemShut
  {NoStop}%
\bibitem [{\citenamefont {Abadi}\ \emph {et~al.}(2016)\citenamefont {Abadi},
  \citenamefont {Barham}, \citenamefont {Chen}, \citenamefont {Chen},
  \citenamefont {Davis}, \citenamefont {Dean}, \citenamefont {Devin},
  \citenamefont {Ghemawat}, \citenamefont {Irving}, \citenamefont {Isard} \emph
  {et~al.}}]{abadi2016tensorflow}%
  \BibitemOpen
  \bibfield  {author} {\bibinfo {author} {\bibfnamefont {M.}~\bibnamefont
  {Abadi}}, \bibinfo {author} {\bibfnamefont {P.}~\bibnamefont {Barham}},
  \bibinfo {author} {\bibfnamefont {J.}~\bibnamefont {Chen}}, \bibinfo {author}
  {\bibfnamefont {Z.}~\bibnamefont {Chen}}, \bibinfo {author} {\bibfnamefont
  {A.}~\bibnamefont {Davis}}, \bibinfo {author} {\bibfnamefont
  {J.}~\bibnamefont {Dean}}, \bibinfo {author} {\bibfnamefont {M.}~\bibnamefont
  {Devin}}, \bibinfo {author} {\bibfnamefont {S.}~\bibnamefont {Ghemawat}},
  \bibinfo {author} {\bibfnamefont {G.}~\bibnamefont {Irving}}, \bibinfo
  {author} {\bibfnamefont {M.}~\bibnamefont {Isard}},  \emph {et~al.},\ }in\
  \href@noop {} {\emph {\bibinfo {booktitle} {OSDI}}},\ Vol.~\bibinfo {volume}
  {16}\ (\bibinfo {year} {2016})\ pp.\ \bibinfo {pages} {265--283}\BibitemShut
  {NoStop}%
\bibitem [{\citenamefont {Yosinski}\ \emph {et~al.}(2014)\citenamefont
  {Yosinski}, \citenamefont {Clune}, \citenamefont {Bengio},\ and\
  \citenamefont {Lipson}}]{yosinski2014transferable}%
  \BibitemOpen
  \bibfield  {author} {\bibinfo {author} {\bibfnamefont {J.}~\bibnamefont
  {Yosinski}}, \bibinfo {author} {\bibfnamefont {J.}~\bibnamefont {Clune}},
  \bibinfo {author} {\bibfnamefont {Y.}~\bibnamefont {Bengio}}, \ and\ \bibinfo
  {author} {\bibfnamefont {H.}~\bibnamefont {Lipson}},\ }in\ \href@noop {}
  {\emph {\bibinfo {booktitle} {Advances in neural information processing
  systems}}}\ (\bibinfo {year} {2014})\ pp.\ \bibinfo {pages}
  {3320--3328}\BibitemShut {NoStop}%
\bibitem [{\citenamefont {Hinton}\ \emph {et~al.}(2012)\citenamefont {Hinton},
  \citenamefont {Srivastava},\ and\ \citenamefont
  {Swersky}}]{hinton2012neural}%
  \BibitemOpen
  \bibfield  {author} {\bibinfo {author} {\bibfnamefont {G.}~\bibnamefont
  {Hinton}}, \bibinfo {author} {\bibfnamefont {N.}~\bibnamefont {Srivastava}},
  \ and\ \bibinfo {author} {\bibfnamefont {K.}~\bibnamefont {Swersky}},\
  }\href@noop {} {\enquote {\bibinfo {title} {Neural networks for machine
  learning lecture 6a overview of mini--batch gradient descent},}\ } (\bibinfo
  {year} {2012})\BibitemShut {NoStop}%
\bibitem [{\citenamefont {Kingma}\ and\ \citenamefont
  {Ba}(2014)}]{kingma2014adam}%
  \BibitemOpen
  \bibfield  {author} {\bibinfo {author} {\bibfnamefont {D.~P.}\ \bibnamefont
  {Kingma}}\ and\ \bibinfo {author} {\bibfnamefont {J.}~\bibnamefont {Ba}},\
  }\href@noop {} {\bibfield  {journal} {\bibinfo  {journal} {arXiv:1412.6980}\
  } (\bibinfo {year} {2014})}\BibitemShut {NoStop}%
\bibitem [{\citenamefont {Hinton}(2012)}]{hinton2012practical}%
  \BibitemOpen
  \bibfield  {author} {\bibinfo {author} {\bibfnamefont {G.~E.}\ \bibnamefont
  {Hinton}},\ }in\ \href@noop {} {\emph {\bibinfo {booktitle} {Neural networks:
  Tricks of the trade}}}\ (\bibinfo  {publisher} {Springer},\ \bibinfo {year}
  {2012})\ pp.\ \bibinfo {pages} {599--619}\BibitemShut {NoStop}%
\bibitem [{Note1()}]{Note1}%
  \BibitemOpen
  \bibinfo {note} {Here we have compared both of the codes with the same set of
  parameters using restricted Boltzmann machine and sampling with the
  Metropolis-Hastings algorithm that flips a random spin. One should notice
  that the NetKet implementation uses a slightly different version of the RBM
  than the one presented here, where the wave function is directly given by
  $\psi (\protect \bm {x},\protect \bm {\theta })=p_{RBM}^m(\protect \bm
  {x};\protect \bm {\theta })$. For a fair comparison of the performance of
  both codes, we used this definition instead of the one presented in Sec.~\ref
  {sec:nnqs} for these tests. The rest of the results were obtained using the
  method presented in Sec.~\ref {sec:nnqs} which allows the use of the Gibbs
  sampling.}\BibitemShut {Stop}%
\bibitem [{Note2()}]{Note2}%
  \BibitemOpen
  \bibinfo {note} {As a reference, at the time of writing, the cost of each
  processor and cost of the graphics processing units is around 500
  USD.}\BibitemShut {Stop}%
\bibitem [{Note3()}]{Note3}%
  \BibitemOpen
  \bibinfo {note} {Here, and for the Heisenberg model discussed in the next
  paragraph, we have used the weights of a system with $64$ spins from the
  respective transfer learning protocol, as in Fig.\ref
  {fig:fig4}.}\BibitemShut {Stop}%
\bibitem [{nsc()}]{nscc}%
  \BibitemOpen
  \href@noop {} {\bibinfo  {journal} {\url{https://www.nscc.sg}}\ }\BibitemShut
  {NoStop}%
\end{thebibliography}%

\end{document}